\documentclass[5p,twocolumn,authoryear]{elsarticle}
\usepackage{longtable}
\usepackage{amssymb}
\usepackage{amsmath}
\usepackage{lineno}
\usepackage[dvipsnames]{xcolor}
\usepackage{booktabs}
\usepackage[T1]{fontenc}
\usepackage{rotating}
\usepackage{caption}
\usepackage{tabularx}
\journal{arXiv}
\usepackage{tikz}
\usetikzlibrary{shapes.geometric, arrows.meta, positioning, calc, fit}

\begin{document}
\begin{frontmatter}
\title{A Scoping Review of the Negative Effects of Digital Technology on Cognition} 

\author[1]{Urška Žnidarič\corref{cor1}} 

\author[1]{Erik Štrumbelj} 

\author[1]{Octavian Machidon} 

\cortext[cor1]{Corresponding author}

\affiliation[1]{organization={University of Ljubljana, Faculty of Computer and Information Science},
            addressline={Večna pot 113}, 
            city={Ljubljana},
            postcode={1000}, 
           country={Slovenia}}

\begin{abstract}
The rapid integration of digital technology into daily life has prompted sustained concern regarding its impact on human cognition. To characterize documented negative effects and the conditions under which they arise, we conducted a scoping review isolating the documented negative effects of digital technology use on cognition. Using a hybrid automated and manual search strategy, we identified foundational seed papers via Scopus and executed an algorithmic citation snowballing process via the OpenAlex API to capture relevant empirical and non-empirical literature. The resulting synthesis of 937 papers (584 empirical, 353 non-empirical) spans legacy screens, multitasking, smartphones, and the nascent work on generative artificial intelligence (AI). Evidence suggests an evolution in the nature of cognitive risk: while research on earlier technologies predominantly describes disruptions to resource allocation, early findings on AI point toward a hypothesized erosion of higher-order cognition. We analyze these risks across cognitive domains through four mechanisms: functional interference, neurochemical dysregulation, structural neuroplasticity, and psychosocial displacement. Effects are frequently moderated by socioeconomic status and environmental factors, suggesting that cognitive decline is often mediated by the displacement of activities rather than direct technological toxicity. Finally, the paper examines how habitual digital offloading could theoretically deplete cognitive reserve, creating downstream risks for long-term health. The collective evidence suggests an efficiency-atrophy paradox, where digital tools optimize short-term performance at the potential expense of long-term unassisted cognition.
\end{abstract}


\end{frontmatter}


\section{Introduction}\label{sec:introduction}

For nearly as long as digital technologies have existed, concerns have been raised about the potential negative effects they may have on users. Academic interest in these issues has been consistent, typically intensifying whenever a new technology emerges. Early academic work on such influential technologies included TV, computers, and telephones. In recent decades, the focus has been on smartphones, social networking platforms, video games, and the internet more broadly.

The latest surge in scholarly and public attention has been driven by artificial intelligence (AI), especially generative AI systems, including conversational assistants, coding assistants, and automated productivity tools. In this paper, we use the more general term AI, while focusing specifically on these types of systems.

AI is particularly interesting because of the extent to which it can perform cognitively intensive tasks, such as researching, writing, coding, visualizing, and reasoning. Its capabilities have advanced to the point that many tasks once considered exclusively human can now be partially automated, something that would have been inconceivable even five years ago. These developments also have the potential to influence aspects of human cognition in ways that have not been observed at this scale, including creativity and critical thinking. However, while the theoretical risks are high, the empirical evidence base for the effect of AI is still in its infancy compared to the decades of research on screen-based media. Consequently, interest in understanding AI’s potential effects on cognition is growing, both within academia and the broader public.

In our view, a key challenge in a technology-driven society has been, and continues to be, optimizing human performance while preserving cognitive integrity. Achieving this goal requires a thorough understanding of how technology affects cognition, enabling awareness of potential trade-offs, and informing strategies to mitigate negative effects. A key question, therefore, is how much we currently understand about the adverse impacts of technology on cognition.

Answering this question is challenging because the field is both broad and fragmented. A wide range of technologies, usage modalities, and target populations has been studied. Moreover, there are many confounders and interactions among different technologies, which often cannot be considered independently. Crucially, many observed negative associations are highly sensitive to socioeconomic and environmental factors, which frequently moderate the impact of technology. Most existing reviews focus on a single cognitive domain, a single type of technology, and a narrowly defined population (see Table~A.1 in Appendix A).

We believe that the field currently lacks a review that combines sufficient breadth and depth. Our goal is to address this gap through a scoping review of relevant empirical and non-empirical studies examining the negative effects of digital technology use on cognition. While digital technologies undoubtedly offer substantial cognitive benefits, conveniences, and enhancements, our objective in this paper is not to evaluate their net cost-benefit. Instead, we adopt an extreme-case sampling approach, focusing specifically on documented negative effects.

While this approach precludes a claim about net effects, it serves two specific analytical purposes. First, by isolating documented harms, we can identify the specific biological and psychosocial mechanisms through which cognitive impairment arises and the conditions under which they arise. Second, because digital technologies do not exist in a vacuum, we must employ a comprehensive analytical scope. By encompassing this wide timeline of overlapping technologies, we aim to characterize a fundamental evolution in the nature of cognitive risk.

\begin{figure*}[!t]
    \centering
    \begin{tikzpicture}[
        box/.style={draw=blue!40!black, rectangle, rounded corners, align=center, text width=6.3cm, minimum height=2.2cm, fill=blue!3, font=\small},
        arrow/.style={->, ultra thick, >=stealth, draw=gray!80}
    ]

    \node[box, text width=8cm] at (0, 0) (sec23) {
        {\bfseries\textcolor{blue!80!black} Sections 2 \& 3}\\[0.1cm]
        \textbf{Methodology \& Empirical Paper Metadata}\\
        \vspace{0.1cm}
        \scriptsize Summarize the literature search protocol and landscape of the empirical papers.
    };

    \node[box] at (-3.8, -3.5) (sec4) {
        {\bfseries\textcolor{blue!80!black} Section 4}\\[0.1cm]
        \textbf{Negative Effects by Cognitive Domain}\\
        \vspace{0.1cm}
        \scriptsize A thematic catalogue of acute impairments, grouped by basic resource allocation deficits and generative/evaluative erosion.
    };

    \node[box] at (3.8, -3.5) (sec5) {
        {\bfseries\textcolor{blue!80!black} Section 5}\\[0.1cm]
        \textbf{Longitudinal Evidence}\\
        \vspace{0.1cm}
        \scriptsize Examines developmental trajectories, brain structure alterations, and the critical mediating role of environmental confounders.
    };

    \node[box] at (-3.8, -7) (sec6) {
        {\bfseries\textcolor{blue!80!black} Section 6}\\[0.1cm]
        \textbf{Mechanisms}\\
        \vspace{0.1cm}
        \scriptsize Synthesizes the overlapping cognitive symptoms into four root biological and psychosocial pathways.
    };

    \node[box] at (3.8, -7) (sec7) {
        {\bfseries\textcolor{blue!80!black} Section 7}\\[0.1cm]
        \textbf{Long-Term Consequences}\\
        \vspace{0.1cm}
        \scriptsize Examines how habitual digital offloading and deskilling could threaten to erode long-term cognitive reserve and cognitive resilience.
    };

    \node[box, text width=8cm] at (0, -10.5) (sec8) {
        {\bfseries\textcolor{blue!80!black} Section 8}\\[0.1cm]
        \textbf{Integrative Synthesis and Discussion}\\
        \vspace{0.1cm}
        \scriptsize Synthesizes the shift from resource allocation to generative erosion, the efficiency-atrophy paradox, and calls to action.
    };

    
    \draw[arrow] (sec23.south) -- (0, -1.75) -| (sec4.north);
    \draw[arrow] (sec23.south) -- (0, -1.75) -| (sec5.north);

    \draw[arrow] (sec4.south) -- (sec6.north);
    \draw[arrow] (sec5.south) -- ++(0, -1.0) -| ([xshift=1.5cm]sec6.north);
    
    \draw[arrow] (sec6.east) -- (sec7.west);

    \draw[arrow] (sec6.south) -- ++(0, -0.65) -| ([xshift=-1.5cm]sec8.north);
    \draw[arrow] (sec7.south) -- ++(0, -0.65) -| ([xshift=1.5cm]sec8.north);

    \end{tikzpicture}
    \caption{Visual roadmap of the manuscript, illustrating the structural flow from methodological foundation through evidentiary cataloging, mechanistic synthesis, and final integration.}
    \label{fig:roadmap}
\end{figure*}

To guide the reader through this review, Figure~\ref{fig:roadmap} provides a visual roadmap of the manuscript. It outlines the structural flow from our methodological foundation and evidentiary cataloging, through the unifying biological mechanisms and long-term health consequences, to our final integrative synthesis.

\section{Scoping Review Methodology}\label{sec:methodology} 

To systematically map the breadth of research on technologically induced cognitive impairments, we conducted a scoping review, documented in accordance with the PRISMA Extension for Scoping Reviews (PRISMA-ScR) guidelines~\cite{tricco2018prisma}. A summary of our study selection process is provided below, while the complete methodological details can be found in Appendix B. No formal review protocol was registered for this study.

To manage the broad and fragmented nature of this literature, our search strategy was divided into two distinct tracks: a \emph{main} set covering general digital technologies (where no publication year limit was applied), and an \emph{AI} set focused on modern AI tools (restricted to publication years 2023 or later).

Initial seed papers were identified via keyword searches in the Scopus database (conducted between July 2024 and August 2025). Scopus was selected due to its comprehensive, multidisciplinary coverage of peer-reviewed literature, making it ideal for identifying foundational review papers. From these seeds, we utilized the OpenAlex~\citep{priem2022openalex} to execute a semi-automated citation snowballing process. The OpenAlex API was chosen for the programmatic, large-scale citation tracking necessary to navigate such a fragmented field. Additionally, a smaller number of highly relevant papers identified opportunistically during the review process or through manual reference checking were included in the final synthesis.

Included literature was required to examine the effects of any digital technology (for example, screens, internet, smartphones, and AI tools) on human cognition. To be included, papers were required to investigate, document, or theorize negative effects, harms, or impairments. Only papers written in English were included.

The search was an iterative hybrid automated and manual screening process. First, the programmatic snowballing automatically filtered the retrieved citation network to retain only articles, reviews, and book chapters that met predefined bibliographic connectivity thresholds. Second, the resulting subsets of highly connected papers underwent manual screening by the authors, who evaluated titles, abstracts, and full texts against the eligibility criteria. Newly identified papers were included as seeds for the next iteration. This loop was repeated until saturation. A final manual post-processing phase was conducted to remove false positives, duplicates across sets, and unretrievable full texts.

Data charting was conducted manually by the authors. For every empirical paper meeting the inclusion criteria, data were extracted and categorized using a standardized charting framework designed to capture the methodological characteristics and contextual focus of each study. Because the primary aim of this scoping review is to synthesize mechanisms of harm and map the breadth of the literature, data extraction focused on categorical study metadata rather than the extraction of continuous effect sizes. The details of the categorization are provided in Appendix B.3.

In accordance with PRISMA-ScR guidelines, a formal critical appraisal or Risk of Bias assessment of the individual studies was not conducted. Because the primary aim of this scoping review is to map the theoretical and mechanistic breadth of the literature rather than to calculate cumulative, clinically actionable effect sizes, evaluating the methodological quality of individual papers fell outside the scope of this synthesis.

Data synthesis was conducted using both quantitative and qualitative approaches. Quantitatively, the extracted empirical metadata were summarized using descriptive statistics. Additionally, a bibliometric citation network analysis was performed to map relationships and clustering among the included literature based on technology type. Qualitatively, a thematic synthesis was employed to aggregate the documented cognitive harms across different domains and to delineate the underlying biological and psychosocial mechanisms of impairment.

\section{Summary of Empirical Paper Metadata}\label{sec:empirical}

\begin{figure*}[tp]
    \centering
    \includegraphics[width=1.0\linewidth]{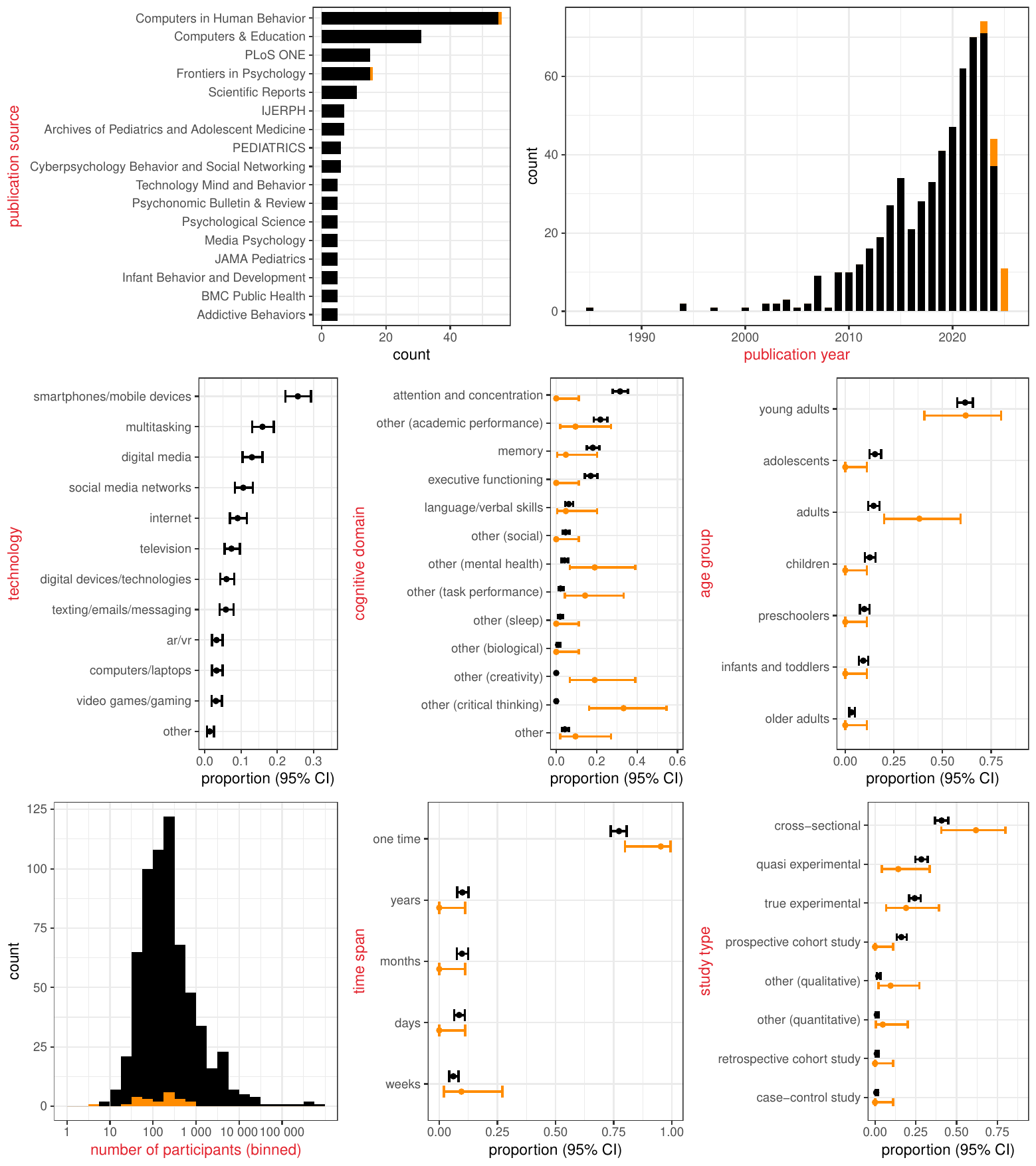}
    \caption{Basic statistical summary of the empirical papers, including a comparison between \textbf{the main group of papers} and more recent \textbf{\textcolor{orange}{papers focusing on AI}}.}
    \label{fig:basic_summary}
\end{figure*}

To contextualize the field, we analyzed the metadata of the 584 empirical papers included in this review, comparing the 563 papers focusing on established digital technologies with the 21 recent papers focusing specifically on AI. A detailed statistical breakdown of these categorizations is provided in Appendix C.

The literature is characterized by rapid growth and high multidisciplinary fragmentation. As shown in the citation network (see Figure~C.3 in Appendix C), research on smartphones, multitasking, and social media forms a well-connected core cluster. The recent AI literature is almost entirely disconnected from this main group, indicating that nascent AI-focused research rarely cites established empirical findings on older technologies.

The data also reveal a stark contrast in research focus (see Figure~\ref{fig:basic_summary}). The main group predominantly investigates attention ($\sim$35\%), memory, and executive functioning, across diverse age groups, especially children and adolescents. Conversely, the AI group targets higher-order cognitive domains, such as critical thinking and creativity, almost exclusively within adult and university-level populations.

Finally, methodological maturity differs significantly between the two eras. While single-time, cross-sectional studies with sample sizes of around 100 participants dominate the entire corpus, the main group contains a much larger proportion of experimental and longitudinal designs. Furthermore, while both groups rely on behavioral and observer-reported assessments, the AI literature currently exhibits an overwhelming reliance on self-reports, reflecting the early, exploratory stage of AI-related behavioral research (see Figure~C.4 in Appendix C).

\section{Negative Effects by Cognitive Domain}\label{sec:negative_effects_domains}

This section serves as a thematic catalogue of the acute and cross-sectional cognitive impairments documented in the literature, organized by specific cognitive domains. Rather than presenting a chronological narrative, this section is designed as a domain-by-domain reference overview. It establishes the evidentiary foundation required to understand the broader risks of digital engagement. 

Because cognitive processes are highly interdependent, readers will note recurring behavioral patterns, such as cognitive offloading or distraction, manifesting across multiple sub-sections. We document these effects here in their specific contexts to provide a comprehensive resource for researchers. However, for an integrative analysis of \textit{how} and \textit{why} these overlapping impairments occur, readers are directed to Section~\ref{sec:mechanisms}.

Given the breadth of the literature, it was not feasible to include all empirical and non-empirical papers identified. This section primarily relies on non-empirical syntheses that aggregate empirical findings, as well as foundational empirical studies. A complete list of the 584 empirical and 353 non-empirical papers identified through the systematic search is provided as Supplementary Material.

\subsection{Part I: Deficits in Basic Resource Allocation}

\subsubsection{Attention Impairment}

Attention-capturing digital environments, characterized by constant notifications, algorithmic feeds, and media multitasking, promote habitual interruption and divided attention, contributing to fragmented attentional systems \citep{firth2024online, barros2024understanding, levine2012mobile}. Experimental, meta-analytic, and review evidence links frequent multitasking and interruption-heavy use to reduced sustained attention, poorer inhibitory control, and learning inefficiencies, with stronger and more consistent effects under high interruption frequency and in younger users \citep{kong2023cognitive, clinton2021stop, eliseev2021externalizing, chen2016does, ershova2019tsifrovoe, liu2017meta}.

Building on this, reviews and meta-analyses show that frequent media multitasking is reliably associated with greater distractibility and poorer sustained attention~\citep{ophir2009cognitive}, particularly in youth. Longitudinal evidence summarized by \cite{alho2022effects} indicates that adolescents with higher media multitasking report increasing attention problems over time, while integrative reviews link multitasking to weaker attention regulation and filtering of irrelevant information \citep{van2015consequences,beuckels2021media}. Meta-analytic syntheses further demonstrate small but consistent associations with everyday attentional failures, and intervention reviews show that reducing multitasking can yield short-term improvements in attention performance, highlighting the sensitivity of sustained attention to interruption-heavy digital habits \citep{parry2021cognitive,wiradhany2021everyday,parry2019media}.

Similarly, the literature on smartphone overuse and early-life screen exposure converges on attentional harms, but the evidentiary base is best read in clusters. For adolescents and adults, systematic and narrative syntheses consistently link smartphone distraction and problematic use to reduced sustained attention and increased distractibility across laboratory and field contexts, with effects moderated by notification frequency, habitual checking, and individual self-regulation~\citep{amalakanti2024impact, taylor2024appetite, liebherr2020smartphones, wilmer2017smartphones, gupta2023unplugged}. These reviews also note overlaps with ADHD-like profiles (inattention, impulsivity, hyperactivity) without making clinical diagnoses, and emphasize that most studies remain correlational and thus limited in establishing directionality~\citep{amalakanti2024impact, liebherr2020smartphones, wilmer2017smartphones}.

For early childhood, the evidence is more explicitly developmental and includes multiple systematic and narrative reviews focused on preschool and infant cohorts. Across studies of children roughly aged 0--6, higher screen exposure is repeatedly associated with weaker sustained attention, greater distractibility, and externalizing behaviors that resemble ADHD-like symptoms, with risk typically increasing at higher doses and in the presence of contextual amplifiers such as sleep disruption, fast-paced content, or limited caregiver scaffolding~\citep{merin2024evaluation, jourdren2023impact, swider2023young, fitzpatrick2023reducing, santos2022association}. Longitudinal findings summarized in these reviews suggest that early and cumulative exposure can predict later attentional difficulties, while also indicating potential bidirectionality~\citep{jourdren2023impact, santos2022association}.

Finally, broader integrative and theoretical reviews situate these associations within behavioral patterns of attentional fragmentation and shallow processing, arguing that technology can drift from a scaffold into a substitute for sustained attentional practice when use is intense, unsupervised, or habitual~\citep{czerniawska2019internet, karkashadze2022digital, harle2012effects, aylsworth2022duty}. 

Research suggests that AI can impair attention by encouraging continuous partial attention and displacing sustained cognitive training. Conceptual and narrative syntheses argue that frictionless, instantly available AI outputs overwhelm limited attentional resources, reduce metacognitive friction, and promote surface-level engagement over prolonged focus \citep{mahajan2025silent, dubey2024redefining, dergaa2024tools}. Empirical classroom evidence further indicates that when AI systems substitute for effortful learning, students show reduced perseverance and deep attentional engagement \citep{ivanov2023dark}.

In educational contexts, algorithmic personalization and rapid content cycles prioritize immediacy over sustained focus, conditioning learners toward fragmented and continuously shifting attention \citep{fasoli2025dark, leon2024potential}. Large-scale narrative reviews consistently link fast-paced, multitask-oriented digital environments to shortened attention spans, habitual task-switching, and weakened attentional persistence \citep{clemente2024digital, de2024understanding}. 

Most observed effects are modest but consistent. Even brief smartphone interruptions impair sustained attention, while heavy use is associated with chronic distractibility~\citep{wilmer2017smartphones, bezgodova2020smartphone, gupta2023unplugged}. Media multitasking shows small-to-moderate negative effects on attention, with stronger impairments when tasks are unrelated, complex, or span different sensory channels~\citep{jeong2016media, chen2016does}. In early childhood, use of mobile devices for more than 2 hours per day is strongly associated with inattention and hyperactivity symptoms \citep{harle2012effects}. While some studies note neutral or context-dependent effects \citep{wiradhany2021everyday, santos2022association}, the overall pattern points to cumulative harm with intensity, content type, and context as key moderators.  

\subsubsection{Memory Impairment}

Research links excessive, fragmented, or externally reliant digital technology use to impairments in memory formation, consolidation, and retrieval. Across smartphones, social media, internet search, and multitasking, heavy or unregulated use consistently weakens internal mnemonic processes and fosters dependence on external aids. Meta-analytic syntheses report small-to-moderate deficits in working and long-term memory under conditions of active smartphone use \citep{bottger2023does, amalakanti2024impact, liebherr2020smartphones}, though recent meta-analyses indicate that the passive "brain drain" effect of mere smartphone presence is highly susceptible to publication bias and often statistically indistinguishable from zero \citep{parry2024does}. Deficits are similarly observed under media multitasking \citep{kong2023cognitive, jeong2016media, beuckels2021media, uncapher2018minds} and habitual online information access \citep{gong2024google, firth2019online, chrzanowski2022changing, czerniawska2019internet}. Narrative reviews further describe convergent neural and behavioral evidence linking chronic digital exposure to reduced activation in neural systems that support memory encoding and retrieval \citep{yuce2025cost, korte2020impact, dikshit2023social, firth2020exploring, avdeeva2022influence}. 

Recent research expands these findings into the era of AI, showing that AI systems amplify the same cognitive offloading~\citep{risko2016cognitive} effects previously observed with search engines and smartphones. Reliance on generative tools such as ChatGPT for information retrieval or task completion has been shown to reduce memory retention and deep encoding, as users engage in passive consumption rather than active rehearsal and consolidation \citep{bai2023chatgpt, dubey2024redefining, paic2025impact, dergaa2024tools, mahajan2025silent}. This continues the well-documented Google effect~\citep{sparrow2011google}, wherein individuals store information externally rather than internally, leading to faster forgetting and weaker recall \citep{gong2024google, eigenauer2024mindware, ali2024understanding, heersmink2024use}.

Beyond declarative memory, AI-assisted technologies are linked to declines in spelling, handwriting, arithmetic, and language learning, as automation reduces the cognitive effort required for these memory-reinforcing skills~\citep{sternberg2024not, jose2025cognitive}. Similar effects are seen with GPS navigation and automated photo documentation, which externalize spatial and episodic memory, weakening hippocampal processing and the vividness of personal recollections \citep{fasoli2025dark, tankelevitch2025tools, grinschgl2022supporting, de2024understanding}. Theoretical and neurocognitive analyses frame this as a progression toward digital amnesia or, a concept popularized by \citet{spitzer2012digitale}, digital dementia~\citep{farkavs2024transforming, clemente2024digital, leon2024potential}. In this view, short-term efficiency comes at the cost of deep learning and durable memory. However, comprehensive meta-reviews frequently challenge these extreme narratives, demonstrating that the aggregate effect of digital media on cognitive performance is often trivial and fears of widespread "digital dementia" are vastly overstated \citep{appel2020are}. Indeed, overall effects are typically small but consistently negative. Light or structured technology use has minimal impact, whereas excessive or unregulated engagement leads to weaker recall, reduced working-memory precision, and slower learning \citep{kong2023cognitive, bottger2023does, amalakanti2024impact, liebherr2020smartphones}. 

\subsubsection{Executive Function Impairment}

Excessive or passive digital engagement is associated with executive-function difficulties across development. In early childhood, heavy or unsupervised screen exposure is linked to poorer attention regulation, working memory, and inhibitory control, largely via overstimulation and displacement of sleep, play, and caregiver interaction \citep{korres2024unsupervised, lillard2015television, anderson2017digital, fitzpatrick2023reducing}. Experimental studies show short-term executive-function depletion following brief exposure to highly stimulating content, while neuroimaging reviews report associations between heavier early exposure and altered neural activation, suggesting vulnerability of executive networks during development \citep{wu2024early, jannesar2023effects, namazi2024immediate}. \citet{lillard2011immediate} demonstrated this acute depletion by showing that just minutes of viewing fast-paced television significantly impaired preschoolers' executive function compared to educational programming or drawing.

In adolescents and adults, executive-function impairment is most consistently linked to heavy media multitasking, problematic smartphone use, and social media addiction. Meta-analyses and systematic reviews report small-to-moderate associations with reduced inhibitory control, working memory, and everyday self-regulation, with effects stronger for self-reported functional difficulties than for performance-based laboratory tasks \citep{kong2023cognitive, parry2021cognitive, wiradhany2021everyday, amalakanti2024impact}. Complementary syntheses indicate that deficits are context-dependent and cue-driven—emerging most clearly under distraction, fatigue, or emotionally salient digital cues \citep{dikshit2023social, cudo2022cognitive, warsaw2021mobile}. 

Overdependence on AI and automation is linked to executive erosion through two recurring pathways: substitution of effortful cognition during task performance, and longer-run deference that weakens autonomy in judgment. In education and knowledge work, instant AI solutions can short-circuit planning, problem-solving, and creative exploration, while inflating perceived competence~\citep{zhai2024effects, shafei2025critical, parsakia2023effect, shanmugasundaram2023impact, dubey2024redefining, ali2024understanding}. Related accounts of AI-enabled cognitive offloading emphasize cumulative deskilling and out-of-the-loop effects, especially for novices and under routine, unscaffolded use~\citep{sutton2018much, poszler2024impact, clemente2024digital}.

A parallel literature focuses on autonomy and decision quality: when algorithmic advice becomes a default, users may defer judgment via automation bias, miscalibrated trust, and responsibility shifting~\citep{steyvers2024three, Eckhardt2024, shukla2025skilling, mahajan2025silent}. These risks motivate calls for human-in-the-loop design and reflective guardrails~\citep{siemens2022human, de2024understanding, paic2025impact, ivanov2023dark, leon2024potential}. Harm is generally defined by prolonged, passive, or high-stimulation engagement, whereas interactive and moderated use appears less detrimental.

\subsection{Part II: The Erosion of Generative and Evaluative Cognition}

\subsubsection{Learning and Academic Performance Impairment}

Extensive empirical evidence indicates that unregulated digital use can impair academic achievement. However, these effects are heavily modality-dependent; major meta-analyses demonstrate that while television viewing negatively impacts academic performance, structured internet and computer use can yield significant positive learning outcomes \citep{marker2018active}. Nonetheless, in unguided contexts, experimental meta-analytic evidence shows that media multitasking produces substantial learning decrements, consistent with divided attention, cognitive overload, and disrupted encoding/comprehension during study and instruction~\citep{jeong2016media, clinton2021stop}.

Across broader higher-education syntheses, the same pattern is repeatedly observed in classrooms and self-study: multitasking with laptops, texting, or social media predicts lower GPA/test performance, poorer note-taking and recall, and reduced learning efficiency~\citep{may2018efficient, beuckels2021media, van2015consequences, carrier2015causes}.

Parallel lines of evidence focus on smartphone use and dependence. Meta-analytic and systematic reviews report small but reliable negative associations between problematic smartphone use and learning outcomes, typically mediated by distraction, time displacement, and downstream pathways such as sleep disruption and stress~\citep{sunday2021effects, amez2020smartphone, ramjan2021negative, gupta2023unplugged}. Social-network use shows a smaller and more heterogeneous relationship with achievement, largely dependent on context (e.g., academically oriented SNS use)~\citep{liu2017meta, doleck2018social, l2013growing, dontre2021influence}. 

A distinct cluster of evidence concerns reading and learning in digital formats. Reviews argue that screen-based and hypertext reading can promote shallow processing and weaker comprehension than print due to higher cognitive load, nonlinear navigation demands, and less accurate metacognitive monitoring~\citep{wylie2018cognitive, lombina2019problemy, baron2021know, skulmowski2023cognitive, chen2016does}. Finally, the literature on immersive environments suggests that immersive VR can increase extraneous cognitive load and impair retention when perceptual richness diverts attention from core material~\citep{poupard2025systematic, zhang2023review, mayer2023promise, hamilton2021immersive}.

Research in AI consistently warns that unreflective reliance can increase short-term efficiency while weakening deep learning and authentic cognitive engagement~\citep{al2024balancing}. Systematic and scoping reviews covering higher-education literatures repeatedly identify cognitive offloading patterns—students using AI to draft, summarize, or solve tasks in ways that reduce independent thinking and encourage surface-level engagement~\citep{rothinam2025systematic, samala2025unveiling, maita2024pros}. Complementing this, a large meta-analysis of ChatGPT interventions finds overall positive learning effects but flags overreliance effects: learning performance tends to decline when use is prolonged or when ChatGPT substitutes for, rather than supports, active processing~\citep{wang2025effect}.

More direct educational syntheses conceptualize this as inversion effects or metacognitive laziness under substitutive AI use~\citep{jose2025cognitive, bauer2025looking, favero2025ai, gerlich2024balancing}. These concerns are echoed in early-childhood and school contexts, which caution that AI can bypass the development and practice of problem-solving and critical thinking skills~\citep{chen2024artificial, singh2024descriptive}.

AI users often demonstrate improved task performance but exhibit weaker engagement and metacognitive awareness, leading to uncritical acceptance and a shallow understanding \citep{krupp2023unreflected,bhuana2025hidden,ccela2024risks}. Several studies highlight a gap between performance and conceptual understanding, showing that AI users often overestimate their competence \citep{fernandes2025ai,krsmanovic2025self}. When used appropriately, however, AI can still provide meaningful learning benefits \citep{fabio2025ai,Alshehri2025ChatGPT}.

Finally, a large conceptual literature argues that these patterns are plausibly self-reinforcing: the speed, fluency, and authoritative tone of AI can normalize shortcut-seeking and reduce tolerance for cognitive effort~\citep{larson2024critical, bai2023chatgpt, burns2025beyond}. Persistent over-reliance is argued to erode the metacognitive foundations necessary for independent decision-making and adaptability~\citep{paic2025impact, duenas2024risks, ivanov2023dark, wiederhold2025cost}. Conversely, educational designs that introduce desirable friction or provocations can preserve user agency, suggesting that harm arises primarily when technology substitutes for, rather than scaffolds, human cognition~\citep{yan2024promises, hou2025appropriate, anna2025extended, yatani2024ai}.

\subsubsection{Critical Thinking Impairment}

Evidence converges that frequent, unguided use of algorithmic systems erodes critical thinking when AI substitutes for human reasoning. Attention-capturing designs and persuasive interfaces promote distraction, heuristic processing, and cognitive offloading, reducing sustained analysis, deliberation, and autonomous judgment~\citep{zimmerman2023attention, vettehen2023attention, talaifar2023freedom, kozyreva2020citizens}.

In education, systematic reviews show that generative AI can support critical thinking under guided, reflective use, but that unstructured or excessive reliance shifts effort from analysis and evaluation to acceptance of AI outputs~\citep{Premkumar2024Impact, maita2024pros, sardi2025generative, Melisa2025CriticalThinkingAI}. Behavioral research further shows that overreliance on authoritative AI advice suppresses reflective evaluation and miscalibrates confidence, leading to deskilled judgment and diminished autonomy~\citep{goyal2025ai, shafei2025critical, Eckhardt2024, klingbeil2024trust}. 

Broader digital-cognition reviews link sustained cognitive offloading and attentional overload to longer-term weakening of executive control and reasoning capacities~\citep{ali2024understanding}. A CHI meta-analysis finds that while AI yields large overall learning gains, these benefits are attenuated or negative for higher-order skills due to offloading \citep{qu2025generative}. Similarly, benefits diminish with prolonged exposure as reliance increases \citep{wang2025effect}. Systematic reviews reinforce this, reporting that unguided or heavy use links to superficial engagement, reduced depth, and dependence \citep{samala2025unveiling, bauer2025looking, shukla2025skilling, hou2025appropriate}.

More targeted syntheses attribute these impairments to cognitive offloading and automation bias: students increasingly outsource analysis, evaluation, and idea generation to AI, gaining short-term fluency while weakening deep understanding \citep{jose2025cognitive, parsakia2023effect, rothinam2025systematic}. Hallucinations and productivity-optimized interfaces further encourage conformity and passive acceptance \citep{chen2024artificial, favero2025ai, mei2025designing}. Meta-analytic and theoretical work additionally shows that these trade-offs persist even when overall performance improves: higher-order reasoning and evaluative depth are limited by homogenization and inflated confidence \citep{alemdag2025effect, holzner2025generativeaicreativitysystematic, grinschgl2022supporting}.

Conceptual and integrative reviews consistently warn that unreflective reliance on AI can produce cumulative deskilling and loss of autonomy \citep{shanmugasundaram2023impact, fasoli2025dark, macnamara2024does, sutton2018much}. Related accounts argue that habitual delegation to AI weakens critical judgment and agency, with risks of broad and intergenerational cognitive atrophy under prolonged, frictionless AI use \citep{kim2025ai, burns2025beyond, peschl2024human, sternberg2024not, mahajan2025silent, leon2024potential}. Drawing on research on transactive memory and the extended mind, these works emphasize that easy, pre-emptive access to answers encourages cognitive miserliness and procedural responding~\citep{atkinson2021hey, czerniawska2019internet, eigenauer2024mindware, bai2023chatgpt, paic2025impact, heersmink2024use}. 

A growing set of AI-focused position papers extends these concerns, arguing that generative systems' fluency and speed amplify dependency and normalize surface-level engagement~\citep{singh2025protecting, wiederhold2025cost, yankouskaya2025can, dergaa2024tools, duenas2024risks, sarkar2024copilot, yatani2024ai, tankelevitch2025tools}. In higher education and knowledge work, unchecked AI reliance is argued to undermine critical thinking and epistemic responsibility at scale~\citep{larson2024critical, ivanov2023dark, singh2024descriptive, anna2025extended,george2024erosion,schenck2024examining}. Users who actively monitor and restrict their AI use demonstrate stronger critical thinking outcomes than passive users \citep{zhou2024mediating}.

\subsubsection{Metacognition Impairment}

Rapidly expanding research indicates that heavy or unreflective AI use erodes metacognitive processes, including self-monitoring, confidence calibration, and reflective evaluation. A recurring behavioral pattern is metacognitive laziness: instant, fluent AI answers reduce the difficulty signals that normally trigger monitoring and regulation, encouraging acceptance over verification---especially for novices~\citep{singh2025protecting, jose2025cognitive, Melisa2025CriticalThinkingAI, rothinam2025systematic}.

A second cluster emphasizes miscalibration under cognitive offloading: relying on external systems can inflate perceived knowledge, increase misattribution of externally provided information to internal competence, and weaken metacognitive sensitivity~\citep{grinschgl2022supporting, mahajan2025silent, duenas2024risks}. In educational contexts, a systematic review links overreliance on AI to reduced self-regulation via automation bias and false confidence from immediate feedback~\citep{goyal2025ai}.

A third cluster frames metacognitive impairment as a workflow shift: when AI generates complete solutions, users move from active creation to critical integration, increasing satisficing (accepting plausible outputs) unless systems deliberately provoke self-critique~\citep{tankelevitch2025tools, sarkar2024copilot, favero2025ai, leon2024potential}. Syntheses highlight poor understanding of AI limits and lack of feedback as drivers of shallow self-evaluation~\citep{steyvers2024three, Eckhardt2024, dumitru2023critical}. Under conditions of routine outsourcing and minimal self-evaluation \citep{shafei2025critical, sardi2025generative}, the literature reports small-to-moderate declines in metacognitive awareness and judgment accuracy \citep{bauer2025looking, yan2024promises, ali2024understanding, fasoli2025dark}.

\subsubsection{Creativity Impairment}

Uncritical AI use is repeatedly linked to dampened originality and weaker divergent thinking, primarily via cognitive offloading and passive acceptance of suggested ideas~\citep{Premkumar2024Impact, samala2025unveiling, yan2024promises}. In educational settings, evidence indicates a fluency–originality trade-off: AI can increase idea output while increasing fixation and lowering creative confidence when users uncritically adopt AI-generated suggestions~\citep{jose2025cognitive}. Complementary design-oriented discussions argue that replacing early-stage ideation with AI reduces practice in exploration and novelty seeking, contributing to longer-run skill erosion~\citep{dubey2024redefining, tankelevitch2025tools}. Broader developmental reviews connect prolonged, passive digital exposure to reduced imaginative flexibility and downstream constraints on creative growth~\citep{ali2024understanding, clemente2024digital}. Theoretical models predict cumulative atrophy and ecosystem-level homogenization through repeated reuse of AI outputs (“knowledge recycling”), producing stagnation in individual and collective creativity~\citep{george2024erosion, peschl2024human, aru2025artificial, leon2024potential}.

Research further indicates that AI influences creativity in both positive and negative ways. On the positive side, AI supports idea generation, collaboration, and creative confidence, particularly in exploratory learning \citep{wei2025effects,habib2024does,lin2024artificial}. At the same time, sustained or unregulated AI use has been associated with reduced originality and homogenized outputs, especially when learners delegate core generative processes to AI rather than engaging in reflective, agentic use \citep{niloy2024chatgpt,mingjie2025double}. Consequences are ultimately dose- and design-dependent, where guided workflows that enforce critique and active interrogation can mitigate harms, transforming AI from a force for homogenization into a scaffold for co-creation.

\subsection{Part III: Broader Psychosocial and Ethical Impacts}

\subsubsection{Emotional and Psychosocial Dysfunction}

Heavy, unregulated, or compulsive digital engagement consistently correlates with anxiety, depression, emotional dysregulation, and lower well-being. Syntheses of youth data show consistent associations between screen activity and internalizing problems, particularly with high social media use and sleep disruption \citep{paulus2023screen, gottschalk2019impacts}. In early childhood, passive exposure without scaffolding predicts socio-emotional delays \citep{massaroni2023relationship, irzalinda2023screen, sadeghi2018impact, uzundaug2022screen}. For adolescents, problematic social networking and multitasking correlate with stress, depression, and impaired self-regulation \citep{hussain2018problematic, carrier2015causes, van2015consequences, warsaw2021mobile, ramjan2021negative}. However, comprehensive umbrella reviews caution that these aggregate associations are typically small, highly inconsistent, and rarely indicative of broad, population-level harm \citep{valkenburg2022social}. Gaming research identifies clinical dysfunction in Internet Gaming Disorder \citep{chitra2023bhushan, ji2022risk}. 

Broader reviews confirm these trends but emphasize that behavioral and emotional risks are often non-linear (Goldilocks effect), where moderate use correlates with optimal psychosocial functioning—a critical buffer for cognitive resilience—compared to total abstinence \citep{przybylski2017digital, ferguson2017everything}, and moderated by context, such as parenting quality and co-viewing \citep{harle2012effects, peric2022pregled, joshi2018information, munsamy2022screen, bulut2023association}.

Recent studies on AI highlight novel psychosocial risks. Over-reliance on chatbots and digital platforms correlates with social isolation, reduced empathy, and the displacement of authentic connection \citep{ali2024understanding, yankouskaya2025can, dergaa2024tools, clemente2024digital, samala2025unveiling, mahajan2025silent}. Interactions with AI tutors can lower self-efficacy and trigger technostress or impostor syndrome \citep{favero2025ai, bauer2025looking, wiederhold2025cost}, while surveillance and automation threaten autonomy and identity \citep{ivanov2023dark, kim2025ai, george2024erosion}. Furthermore, generative AI may undermine emotion regulation and inner speech \citep{leon2024potential, siemens2022human, sternberg2024not}, and dependence on these systems for validation risks diminishing confidence in human judgment \citep{parsakia2023effect}. Multiple studies also document psychological and emotional consequences associated with AI use, linking adoption to increased anxiety, technostress, and symptoms of dependency \citep{litan2025mental,frenkenberg2025s}. AI-based learning environments may additionally increase cognitive load and anxiety, highlighting an important trade-off between performance gains and emotional well-being \citep{khan2025psychological,maral2025problematic}.

\subsubsection{Moral and Ethical Deskilling}

Some works argue that pervasive, unregulated, or outcome-oriented uses of digital technologies can erode users' moral attention, judgment, agency, and responsibility. Foundational virtue-ethical and philosophical accounts frame this as moral deskilling: technologies displace opportunities to practice moral perception, practical wisdom, and responsibility through automation, attention capture, and deference to algorithmic judgment \citep{vallor2015moral, vallor2013future, eigenauer2024mindware}. Related analyses emphasize attentional capture, automation bias, and conformity pressures as mechanisms that weaken ethical agency and accountability in everyday digital environments \citep{zimmerman2023attention, schuster2025attention, kim2025ai, sternberg2024not}.

More recent conceptual syntheses extend these concerns to generative AI, arguing that cognitive offloading, illusion of competence, authorship ambiguity, and normalized deference to machine outputs foster moral abdication and erosion of ethical self-regulation, particularly in education and knowledge work \citep{mahajan2025silent, samala2025unveiling, skulmowski2024placebo}. Overall, the literature is currently dominated by conceptual and narrative reviews, with a smaller number of systematic syntheses \citep{maita2024pros, poszler2024impact}. A quasi-experimental classroom study finds a modest path from AI to development, consistent with displacement of human interaction and reflective work \citep{ivanov2023dark}.


\section{Longitudinal Evidence}\label{sec:longitudinal}

Fully understanding the cognitive consequences of technology use requires evidence that captures how exposure and outcomes evolve over time. Longitudinal and long-term studies, by tracking individuals across meaningful time spans, are uniquely positioned to examine temporal ordering, assess persistence versus transience of effects, and examine whether early exposure alters subsequent cognitive development. 

The vast majority of the long-term evidence focuses on infants and preschoolers, where the technology of interest is predominantly television or generalized screen time. In contrast, the subset of studies involving adolescents and students shifts focus toward interactive technologies, specifically social media, video games, and smartphones. A critical methodological dimension distinguishing these groups is granularity: while early childhood studies typically rely on aggregate duration (total hours/day), adolescent studies are more likely to isolate contextual usage (multitasking, in-class use) to study specific behavioral outcomes.

\subsection{Language, Intelligence, and General Cognitive Development}

In early childhood, higher screen time is frequently associated with delays in language and general cognitive milestones. \cite{Madigan2019-dz} identified a directional association where screen time at 24 and 36 months predicted poorer performance on developmental screening tests later, with no evidence of reverse causality. Similarly, \cite{Aishworiya2019-rs} reported that for every extra hour of TV watched daily at 12 months, composite IQ at age 4.5 decreased significantly. Earlier onset and intensity appear critical, as \cite{Al_Hosani2023-dq} linked device ownership and use between 12 and 24 months to delayed speech, while \cite{Tomopoulos2010-ki} and \cite{Supanitayanon2020-yv} found that media exposure as early as 6 months—particularly content lacking verbal interaction—predicted lower cognitive scores. Trajectory analyses provide further nuance, as \cite{Stockdale2022-ce} identified a high and steady viewing group that performed most poorly on language and executive function at age 4. 

However, effect sizes often attenuate after adjusting for environmental factors, and the direction of the effect may depend on the child's age. \cite{zimmerman2005children} found that while television viewing before age 3 had adverse effects on reading and short-term memory, viewing between ages 3 and 5 actually had a modest beneficial effect on those same domains. Furthermore, \cite{Martinot2021-co}, \cite{yang2024association2}, and \cite{yang2024associations} all found that total screen duration frequently showed weak or non-significant associations with verbal IQ or intellectual abilities after adjustment, but the context of exposure—specifically TV during family meals—consistently predicted lower scores. Socioeconomic status (SES) also acts as a powerful moderator. \cite{Slobodin2024-ef} found screen time predicted language deficits primarily in moderate-high SES families, while \cite{Kuhhirt2020-be} found no substantive associations with cognitive ability after adjusting for confounders.

The impact of technology on intelligence extends into adulthood and across infrastructure shifts. \cite{ackermann2023broadband} found that the expansion of high-speed internet access resulted in a decline in crystallized and fluid intelligence among young adults. Conversely, content type remains a key factor in older children. \cite{Sauce2022-hm} found that while TV was negatively correlated with intelligence, video gaming actually positively impacted the amount of change in intelligence over a two-year period. This aligns with foundational meta-analyses demonstrating that active engagement, such as playing action video games, reliably enhances top-down attention and spatial cognition \citep{bediou2018meta}. Across the lifespan, heavy use continues to correlate with cognitive decline. \cite{Fancourt2019-fy} observed a dose-response decline in verbal memory among adults over 50 watching more than 3.5 hours of TV per day. Finally, the role of displacement is highlighted by \cite{Sugiyama2023-ec}, who found that the negative impact of screens on daily living skills was partially mediated (18\%) by the frequency of outdoor play.

\subsection{Attention, ADHD, and Behavioral Problems}

A significant body of longitudinal research connects early screen exposure to later attentional deficits and behavioral issues. \cite{Christakis2004-gn} found that every hour of daily TV at ages 1 and 3 increased the risk of attentional problems at age 7 by 9\%, a finding echoed by \cite{Tamana2019-pg}, who reported that preschoolers watching more than 2 hours daily were nearly 8 times more likely to meet ADHD criteria. \cite{zimmerman2007associations} specified that entertainment content before age 3—but not educational content—predicted these problems. Broader behavioral risks also emerge, as \cite{Chen2020-xx} linked early screen exposure to autistic-like behaviors, while \cite{Srisinghasongkram2021-he} found that screen media multitasking at 18 months predicted behavioral problems at ages 4 and 6.

These associations persist from childhood into early adulthood. \cite{Landhuis2007-ha} and \cite{Soares2022-mj} found that childhood or adolescent television viewing predicted attention problems and ADHD diagnoses. In adolescence, the focus shifts to interactive digital media. \cite{ra2018association}, \cite{Boer2020-sk}, and \cite{wallace2023screen} all linked high-frequency social media or digital media use to subsequent ADHD symptoms, with \cite{wallace2023screen} identifying impulsivity as a key mediator. Gender differences may also exist, as \cite{George2021-dn} found frequent texting was associated with inattention, specifically in girls.

The directionality of these effects is increasingly examined. While \cite{Hill2020-ou} suggested that children with existing ADHD concerns are given more screen time, \cite{Nagata2024-wt}, \cite{Almeida2023-dm}, and \cite{Sihoe2023-gi} provide prospective evidence that screen time or smartphone addiction precedes the development of inattention and symptoms of mental illness. Notably, \cite{Shih2023-yj} identified a unique pathway: maternal screen time of more than 3 hours/day when the child was 3 was a significant predictor of ADHD by age 8, suggesting that parental habits may be as influential as the child's own exposure. Finally, the negative association between technology and well-being appears small in large-scale cross-sectional data \citep{Orben2019-ag}, though sleep disruption remains a reliable longitudinal outcome of bedtime use \citep{Orben2020-oj}.

\subsection{Executive Function and Brain Structure}

Excessive screen exposure is robustly associated with deficits in executive function and measurable changes in brain development. \cite{barr2010infant} found that high exposure to adult-directed TV in infancy predicted poorer executive function and lower IQ at age 4. \cite{Corkin2021-fu} added that TV exposure during meals at age 2 was specifically linked to poorer performance on hot executive tasks. These difficulties may be more acute for specific populations. \cite{Oh2023-bl} demonstrated that children with lower baseline IQ suffered significantly more severe executive function  difficulties from high media exposure than their higher-IQ peers, while \cite{vohr2021association} linked screen time in extremely preterm children to inhibition and metacognition deficits. Laboratory results are further complicated by context, as \cite{Portugal2023-my} found that while high touchscreen use was associated with reduced working memory concurrently, these effects became non-significant when controlling for background TV viewing.

Neuroimaging and EEG data provide a physical timeline for these behavioral findings. \cite{Law2023-qs} found that screen time at 12 months was associated with altered cortical EEG activity (higher theta/beta ratio) at 18 months, which partially mediated later executive function impairments. Regarding brain structure, \cite{Takeuchi2016-fv} and \cite{Takeuchi2018-ve} found that frequent internet and video game use were associated with smaller developmental increases in regional gray and white matter volume and delayed microstructural development in the basal ganglia and prefrontal cortex. Furthermore, \cite{He2023-uc} linked social media use to reduced volume in the hippocampus and caudate, while \cite{Song2023-au} observed that a video-centric media pattern was associated with persistent differences in functional connectivity within attention and default mode networks.

\subsection{Long-term Academic and Educational Outcomes}

The functional consequences of technology-induced cognitive interference manifest in long-term academic trajectories. \cite{vanderloo2022children} found that early childhood screen use was longitudinally associated with increased vulnerability in school readiness. In primary education, \cite{Rosen2016-dt} and \cite{Dempsey2019-qe} identified weekly computer use and mobile phone ownership by age 9 as significant negative predictors of reading and math achievement. \cite{Luo2020-cu} further established that media multitasking negatively predicted academic performance over a 6-month period.

Policy interventions suggest these trends are reversible, as \cite{beneito2022banning} found that banning mobile phones in school significantly improved PISA scores, equivalent to nearly a year of learning in science. However, without such interventions, negative impacts persist into higher education. \cite{Amez2023-fe} and \cite{Bjornsen2015-px} identified a causal negative relationship between smartphone/social media use during class and average exam scores. These academic gaps may have lifetime consequences, as \cite{Johnson2007-oh} followed youths for nearly two decades and found that those watching $\geq$3 hours of TV daily during adolescence were significantly less likely to obtain postsecondary education by age 33.

\subsection{Other Relevant Longitudinal Findings}

Beyond the core cognitive domains, some studies offer insights into broader developmental outcomes. \cite{vanderloo2022children} found that increased daily screen use was longitudinally associated with reduced language and cognitive development domain scores. \cite{Sugiyama2023-ec} observed that higher screen time (more than 1 hour/day) at age 2 was significantly associated with lower communication scores and daily living skills at age 4, with the latter partially mediated by reduced outdoor play. \cite{Supanitayanon2020-yv} highlighted that the lack of verbal interaction during screen use in the first two years predicted lower cognitive development at age 2, which subsequently predicted lower cognition at age 4.

\subsection{Overview of Common Control Variables}
\label{sec:controls}

Longitudinal studies examining the relationship between technology use and cognitive development employ a broad set of covariates to address confounding. These controls are typically chosen based on the developmental stage of the study population.

Across all age groups, the most consistently included controls are demographic characteristics and SES. In studies of children and adolescents, SES is commonly operationalized using household income, poverty status, and parental education, with maternal education frequently highlighted as a key predictor~\citep{Aishworiya2019-rs,Kuhhirt2020-be,vohr2021association}. In adult samples, SES is usually measured by participants’ educational attainment, employment status, and wealth~\citep{ackermann2023broadband,Fancourt2019-fy}. Age, sex or gender, and race or ethnicity are routinely controlled for across populations~\citep{barr2010infant,Christakis2004-gn}. Family structure variables, including marital status, single-parent households, and number of siblings, are also commonly included~\citep{Amez2023-fe,Madigan2019-dz}.

Many studies additionally control for prior ability, environmental factors, and behavior. To address reverse causality, baseline measures of IQ, executive function, or developmental scores are often included~\citep{Landhuis2007-ha,Oh2023-bl,Shih2023-yj}. In student samples, prior grades or standardized test scores are used~\citep{Amez2023-fe,Bjornsen2015-px,Dempsey2019-qe}. Lifestyle factors such as sleep duration, physical activity, and substance use are frequently measured~\citep{Tamana2019-pg,Nagata2024-wt,Orben2019-ag}. In children, studies often adjust for cognitive stimulation in the home, maternal mental health, and parenting style~\citep{Christakis2004-gn,Tomopoulos2010-ki,Supanitayanon2020-yv}. In adolescent and adult samples, psychological traits such as personality, self-control, and motivation are commonly included~\citep{Bjerre-Nielsen2020-wl,ackermann2023broadband}.

Several variables consistently account for a substantial share of the observed associations between technology use and negative cognitive outcomes. In child studies, parental education and SES emerge as the strongest confounders, as lower parental education predicts both higher screen exposure and lower cognitive scores. Adjusting for these factors often markedly attenuates negative estimates~\citep{Kuhhirt2020-be,vohr2021association}. Prominent reviews further argue that widespread "tech panics" are largely unsupported by aggregate data, noting that negative associations are often trivial in size and heavily confounded by offline vulnerabilities that simply mirror online \citep{odgers2020annual}. Similarly, when time-invariant individual characteristics (genetics, intelligence, or personality) are accounted for using fixed-effects models, associations typically shrink, suggesting that stable traits drive both media use and academic outcomes~\citep{Bjerre-Nielsen2020-wl}. Controlling for prior behavioral problems frequently reveals evidence of reverse causality, as children with attention or behavioral difficulties are more likely to be exposed to screens~\citep{Landhuis2007-ha}. Contextual factors, such as background television or co-viewing, explain outcomes more effectively than total screen duration~\citep{Martinot2021-co,Portugal2023-my}. Other examples of studies where no effect is found when accounting for confounders: Associations between screen exposure and motor deficits disappeared after controlling for SES and gender~\citep{Slobodin2024-ef}, and total daily screen time was not associated with verbal IQ once home environment and parental education are included~\citep{Martinot2021-co}. Similarly, screen use trajectories were not associated with later intellectual abilities after adjustment for sociodemographic characteristics, baseline cognition, and lifestyle factors~\citep{yang2024association2}, and daily screen time at age 3.5 shows no association with general cognitive development at age 5.5 in fully adjusted models~\citep{yang2024association}.

\section{Mechanisms of Technologically-Induced Cognitive Impairment}\label{sec:mechanisms}

The cognitive impairments detailed in Sections~\ref{sec:negative_effects_domains} and \ref{sec:longitudinal} such as deficits in memory, critical thinking, and creativity, do not occur in isolation. Because cognitive processes are highly interdependent, these impairments are driven by shared behavioral root causes. For example, cognitive offloading acts as a unifying behavioral driver: when a user substitutes technology for effortful thought, it simultaneously degrades memory consolidation, diminishes metacognitive monitoring, and bypasses creative generation. 

To synthesize this overlap and provide an integrative framework, we categorize the root mechanisms driving these impairments into four pathways (summarized in Table~\ref{tab:mechanisms}): \textit{functional interference} (real-time processing bottlenecks); \textit{neurochemical and metabolic dysregulation} (disruption of homeostatic signaling); \textit{structural neuroplasticity} (morphological changes due to disuse); and \textit{psychosocial and developmental displacement} (indirect costs via the usurpation of critical activities). We further distinguish between acute effects (immediate functional disruptions) and chronic effects (long-term structural adaptations).

Note that while experimental, neuroimaging, and longitudinal studies provide converging evidence of functional and structural differences associated with patterns of digital technology use, direct longitudinal evidence demonstrating progressive within-individual neural change attributable to technology exposure remains limited. Importantly, much of the available neurobiological evidence derives from task-based or cross-sectional neuroimaging paradigms, which identify correlational activation differences rather than demonstrating stable or experience-dependent neural reorganization in naturalistic settings. Crucially, the mechanisms outlined below are highly sensitive to environmental moderators; as established in Section \ref{sec:longitudinal}, the magnitude of these effects often depends on the presence of parental scaffolding and socioeconomic stability. Accordingly, these pathways illustrate plausible processes through which observed cognitive effects may arise, rather than definitive causal explanations.

\begin{table*}[ht]
\centering
\caption{Categorization of mechanisms of technologically-induced cognitive impairment.}
\label{tab:mechanisms}
\begin{tabular}{p{5cm} p{12cm}}
\hline
\textbf{Category} & \textbf{Primary Manifestations}  \\
\hline
Functional Interference & Cognitive overload, attention residue, DLPFC bottlenecking, failure of memory encoding, acute automation bias. \\
Neurochemical \& Metabolic Dysregulation & Dopaminergic overstimulation, reward deficiency syndrome, melatonin suppression, technostress (cortisol). \\
Structural Neuroplasticity & Functional weakening of ACC or Hippocampus, changes in gray matter density, synaptic pruning, neuronal recycling. \\
Psychosocial \& Developmental Displacement & Loss of sleep, physical activity, family interaction, productive struggle, and waking rest. \\
\hline
\end{tabular}
\end{table*}

\subsection{Functional Interference}

Functional interference refers to acute bottlenecks in processing capacity that occur during real-time usage. These mechanisms primarily impair attention, memory encoding, and executive oversight.

\subsubsection{Cognitive Overload and Gating}

Cognitive overload occurs when high extraneous load, driven by multitasking, rapid notifications, or complex AI outputs, exceeds the capacity of the Dorsolateral Prefrontal Cortex (DLPFC) to process information. When the DLPFC reaches its limit, the Prefrontal Cortex (PFC) gates or rejects incoming sensory information. This results in a functional blindness to new inputs, severely impairing memory encoding and learning comprehension. This effect was fundamentally demonstrated by \citet{strayer2003cell}, who showed that cognitive distraction from cell phone use causes inattentional blindness, severely impairing the encoding of visual information even when eyes remain on the primary task.

At the level of core mechanisms, large-scale syntheses identify capacity depletion as the dominant pathway: concurrent media use overtaxes limited attentional and working-memory resources, weakening executive control and reducing the efficiency of encoding and retrieval~\citep{beuckels2021media, parry2021cognitive}. Experimental and classroom studies specify interference and task-switching costs as proximate mechanisms, whereby notifications, texting, or social media interrupt attentional shielding, force costly reorientation, and fragment ongoing cognitive processing, leading to poorer recall and comprehension~\citep{poplawska2021we, jeong2016media, chen2016does}. Persistent exposure to these fragmented states provides a mechanistic basis for the longitudinal ADHD trajectories described in Section 4.2, suggesting that chronic divided attention may functionally train the brain for distractibility.

This cognitive depletion aligns with the brain drain hypothesis, which posits that the mere physical presence of a smartphone reduces available working memory capacity, as individuals must continuously expend cognitive resources to inhibit automatic orientations toward their device \citep{misra2016iphone,ward2017brain}.

In multimedia learning contexts, cognitive load theory provides a complementary mechanism: extraneous stimuli compete with germane processing, slowing reading and degrading comprehension accuracy when pacing control is low or concurrent channels draw on shared resources~\citep{zhang2023review, poupard2025systematic, mayer2023promise}.

\subsubsection{Attention Residue}

Rapid task switching triggers attention residue, a failure in cognitive transition that implicates the Anterior Cingulate Cortex (ACC). Under conditions of rapid switching, the ACC fails to fully inhibit the neural networks associated with the initial task before activating those required for the subsequent one~\citep{mark2023attention, uncapher2018minds}. This incomplete inhibition leaves cognitive resources partially occupied by the previous context, degrading performance on the current task and preventing deep focus.

\subsubsection{Acute Automation Bias}

In the context of AI, functional interference manifests as acute automation bias. The brain downregulates activity in the ACC and DLPFC when presented with authoritative automated outputs. This bypasses the inhibitory control required for verification, leading to the reflexive acceptance of algorithmic suggestions and a failure of critical thinking and executive oversight~\citep{poszler2024impact, sutton2018much, zhai2024effects}. While Section 3 establishes this bias in acute tasks, its longitudinal impact on generative capabilities remains a theoretically grounded concern awaiting cohort validation.

\subsection{Neurochemical and Metabolic Dysregulation}

This category encompasses the disruption of homeostatic signaling systems, including neurotransmitters (dopamine), hormones (cortisol, melatonin), and metabolic regulation. These mechanisms primarily impair inhibitory control, emotional regulation, and memory consolidation.

\subsubsection{Dopaminergic Overstimulation and Reward Conditioning}

A primary driver of attentional failure is the disruption of top-down control via the Dopaminergic System (DS). Adhering to an Inverted-U dose-response curve, the excessive phasic spikes triggered by high-intensity digital stimuli flood D1 receptors in the PFC. This hyper-stimulation suppresses neuronal firing in the PFC, shifting neural control from the reflective PFC to the reflexive Striatum~\citep{weinstein2015new, weinstein2017new, weinstein2020neurobiological, yankouskaya2025can, li2020relationships}. This shift impairs volitional intent and promotes impulsive behavior.

Chronically, this leads to a reward deficiency syndrome and conditioning: Sustained overstimulation causes receptor downregulation (tolerance). As sensitivity drops, the brain requires increasingly intense stimuli to maintain arousal, rendering low-intensity tasks (for example, reading) unrewarding and driving compulsive consumption~\citep{ali2024understanding, firth2019online, lissak2018adverse}. Persuasive design features (variable rewards, infinite scroll) condition the brain to seek the act of task-switching itself as a reward. Since novelty triggers dopamine, the act of shifting attention is reinforced, training the brain to habitually reject sustained focus~\citep{haidt2024anxious, vettehen2023attention, kozyreva2020citizens}.

\subsubsection{Circadian Rhythm Disruption}

Exposure to blue light suppresses melatonin production, delaying circadian rhythms and impairing sleep quality. This disruption compromises the restorative processes essential for memory consolidation and emotional regulation, particularly in adolescents~\citep{amalakanti2024impact, goldon2024short, butler2024critical, virgilio2024cognition, gottschalk2019impacts}. As demonstrated in the longitudinal analysis by \cite{Chen2020-xx}, this mechanism acts as a critical mediator between early screen exposure and later behavioral problems, framing sleep loss as a primary physiological pathway for cognitive decline.

\subsubsection{Technostress and Physiological Arousal}

The sheer intensity and perpetual novelty of digital interfaces contribute to technostress, characterized by physiological hyperarousal (elevated cortisol, chronic low-grade inflammation) and psychological burnout. This state taxes regulatory capacities beyond simple sensory arousal, leading to anxiety, fear of missing out, and reduced affect regulation~\citep{pothuganti2024technostress, limone2021psychological, hussain2018problematic, ramjan2021negative}.

\subsection{Structural Neuroplasticity}

Chronic functional disengagement leads to physical morphological changes via the use-it-or-lose-it principle. These mechanisms primarily affect long-term memory, sustained attention, and creative synthesis.

\subsubsection{Associated Structural Changes}

While establishing definitive causality remains difficult, sustained reliance on bottom-up stimulation rather than top-down regulation is associated with structural deficits. Multitaskers exhibit reduced gray matter volume in the ACC and weakened functional connectivity between the ACC and Precuneus, a circuit vital for self-awareness and focus~\citep{alho2022effects, firth2020exploring, yuce2025cost}. Similarly, neuroimaging documents cortical thinning in the PFC and orbitofrontal cortex in heavy users, correlating with impaired executive control and emotion regulation~\citep{korte2020impact, paulus2023screen, collerone2013cervello}. These neuroimaging findings are corroborated by longitudinal studies~\citep{Takeuchi2018-ve}, linking frequent internet use to smaller developmental increases in brain volume related to language and executive function.

\subsubsection{The Google Effect and Memory Systems}

The Google Effect refers to cognitive offloading of information retrieval, which diminishes hippocampal engagement. Chronic reliance on external storage leads to synaptic pruning of underutilized memory networks, potentially driving digital amnesia~\citep{wiederhold2025cost, risko2016cognitive, grinschgl2022supporting}. Furthermore, divided attention shifts activity to the Striatum, resulting in observed striatal hypertrophy and the formation of rigid, habit-based memories rather than flexible, declarative ones~\citep{chen2016does, roussos2023mind}.

\subsubsection{Neuronal Recycling and Creativity}

Theoretical neuroscience suggests that persistent offloading of creative tasks to AI may reconfigure neural niches. Without the cognitive friction of independent problem-solving, the brain ceases to recruit specific cortical territories (for example, PFC-DMN loops). These substrates may be repurposed for passive discrimination rather than creative synthesis, effectively eroding the biological machinery required for deep, autonomous cognition~\citep{ali2024understanding, clemente2024digital, george2024erosion, leon2024potential, aru2025artificial}.

\subsection{Psychosocial and Developmental Displacement}

This mechanism serves as the primary scientific bridge between the thematic harms of Section 3 and the messy, confounder-sensitive data of Section 4. Displacement suggests that technology's negative effects are often indirect results of the loss of essential biological and social maintenance activities.

\subsubsection{Developmental Scaffolding}

In early childhood, passive screen time displaces caregiver interaction, physical play, and language-rich engagement. This displacement correlates with delays in language acquisition, motor development, and executive function, as the developing brain is deprived of necessary social and physical inputs~\citep{haidt2024anxious, xie2024screen, fitzpatrick2023reducing, madigan2020associations, anderson2017digital}. This is consistent with Section 4.1 findings where screen duration effects often disappear when family mealtime interaction and high-quality scaffolding are maintained.

\subsubsection{The Productive Struggle of Learning}

Generative AI displaces the productive struggle necessary for expertise. By offering instant outputs, tools like ChatGPT short-circuit the neural consolidation of skills such as writing, coding, and problem-solving. This displaces the deliberate practice required for skill formation and higher-order reasoning~\citep{singh2025protecting, burns2025beyond, bai2023chatgpt, sternberg2024not, rothinam2025systematic}.

\subsubsection{Social and Moral Practice}

Machine-mediated interactions displace authentic human connection, fostering parasocial dependence and reducing empathy and theory of mind capacities~\citep{turkle2011alone, kim2025ai, parsakia2023effect, samala2025unveiling}. Furthermore, the automation of choices removes opportunities to exercise moral perception, leading to the atrophy of ethical reasoning and caring skills due to a lack of practice~\citep{vallor2015moral, vallor2013future, eigenauer2024mindware, mahajan2025silent}.

\subsubsection{Waking Rest and Physical Activity}

Finally, continuous digital consumption displaces micro-rests, moments of waking idleness mediated by the Default Mode Network. These moments are essential for neural replay and memory consolidation. Their loss effectively blocks the stabilization of recent memories~\citep{immordino2012rest}. Furthermore, as demonstrated in Section 4.1, the negative impact of screens is significantly mediated by the displacement of outdoor play (\cite{Sugiyama2023-ec}), highlighting how lifestyle substitution explains the variance often attributed to screen duration.

\subsection{Systemic Negative Feedback Loops}

A critical finding across these mechanisms is that they rarely operate in isolation. Rather, they form recursive, self-reinforcing cycles where the cognitive impairment itself drives increased reliance on the technology. We identify three primary negative feedback loops that perpetuate excessive use and dependency.

\subsubsection{The Neurochemical Loop: Tolerance and Compulsion}

The most fundamental loop is driven by the downregulation of dopaminergic receptors. As chronic overstimulation reduces receptor sensitivity, the user experiences a drop in baseline arousal and satisfaction. This tolerance renders low-stimulation activities unrewarding, driving the user to seek increasingly intense digital stimuli to reach the same hedonic threshold. Consequently, the mechanism of impairment directly necessitates increased consumption, shifting behavior from voluntary engagement to compulsive regulation of neurochemistry.

\subsubsection{The Cognitive Loop: Weakening and Dependency}

In the domain of AI and executive function, cognitive offloading creates a cycle of deskilling and dependency. As users rely on external tools for memory retrieval or reasoning, their internal capabilities for these tasks may weaken due to disuse. This reduction in internal competence lowers the user's cognitive self-efficacy and confidence, making them less capable of performing the task without assistance in the future. The result is a dependency spiral: offloading leads to skill erosion, which necessitates further offloading, transforming a temporary scaffold into a permanent crutch.

\subsubsection{The Psychophysiological Loop: Fatigue and Disinhibition}

Finally, the disruption of sleep and circadian rhythms creates a physiological loop of disinhibition, leading to cumulative sleep debt and daytime fatigue. Since the PFC is highly sensitive to metabolic fatigue, sleep deprivation disproportionately impairs top-down inhibitory control. A fatigued brain is less able to resist bottom-up sensory drives, leading to increased passive screen time and sedentary behavior, which further disrupts sleep quality. Thus, the physiological consequence of overuse erodes the inhibition processes required to regulate that use.

\section{Long-term Consequences: Cognitive Epidemiology and Reserve}\label{sec:health}

Having established that unregulated digital engagement can impair cognition, it is critical to contextualize the broader stakes of this decline. The field of cognitive epidemiology positions early-life cognitive ability as a robust predictor of longevity and physical health~\citep{calvin2011intelligence, martin2004cognitive}. Consequently, any environmental factor, such as pervasive digital offloading, that suppresses cognitive development or accelerates atrophy may carry downstream risks for chronic disease and mortality.

\subsection{Cognition as a Determinant of Physical Health}

Lower cognitive ability, particularly intelligence (IQ), is consistently associated with reduced longevity and a wide range of chronic conditions \citep{gottfredson2004intelligence, fries2022intelligent}. Specifically, a 15-point IQ disadvantage in early life has been linked to a 22\% higher risk of subsequent illness \citep{fries2025multilevel}, including cardiovascular diseases (stroke, myocardial infarction) \citep{kaisaier2025causal, yang2022low, bardugo2024cognitive}, diabetes, and arthritis \citep{wraw2015intelligence}. Poor cognition also correlates with adverse oral health \citep{abramovitz2021cognitive} and an increased risk of severe mental disorders, notably depression and schizophrenia \citep{osler2007cognitive, christensen2018young}. Conversely, strong memory performance buffers against physical health decline in later life \citep{nelson2020bidirectional}.

This evidence base rests on extensive systematic reviews and longitudinal cohorts \citep{batty2007premorbid, fang2024association, sanchez2023intelligence, dobson2017associations, li2024association, zhao2025role, yu2025adverse, ball2024childhood}. While some older cohorts lacked female representation, the overarching pattern is robust across diverse populations \citep{whalley2001longitudinal, hemmingsson2006association, twig2018cognitive, batty2008does, der2009association, hart2004childhood, hemmingsson2009cognitive, kajantie2012stroke, lawlor2008association, batty2009iq, vcukic2017childhood, jokela2011sibling, murray2011association, stilley2010impact, khandaker2018childhood, whitley2010intelligence, schmidt2013cognitive}.

The mechanisms linking cognition to health are multifaceted \citep{deary2021intelligence}:
\begin{itemize}
    \item \textbf{Socioeconomic Mediation:} Intelligence correlates with education and employment, enhancing access to resources that protect health \citep{davies2019multivariable, howe2023educational}.
    \item \textbf{Health Literacy and Adherence:} Higher cognitive ability facilitates the understanding of complex health instructions and adherence to preventive treatments \citep{serper2014health, zullig2014health}.
    \item \textbf{System Integrity:} A system integrity hypothesis suggests that overlapping genetic factors influence both neural efficiency and bodily health, explaining associations with obesity and cardiovascular resilience \citep{deary2019genome, hagenaars2017cognitive, yang2022low}.
    \item \textbf{Behavioral Regulation:} Higher cognition correlates with reduced risks of smoking and alcohol dependence, likely due to superior inhibitory control \citep{li2023can, gage2018investigating}.
\end{itemize}

\subsection{Cognitive Reserve and Digital Atrophy}

The most direct implication for the digital age involves Cognitive Reserve (CR), the brain's capacity to improvise and recruit alternative networks to cope with pathology \citep{stern2002cognitive, guan2024linking}. CR posits that individuals with robust neural networks, built through lifelong education, complex work, and cognitive stimulation, can better withstand the physical impacts of aging and disease \citep{dove2024high, stern2012cognitive, tomaszewski2018compensation}.

Higher CR significantly lowers the odds of clinical dementia and mild cognitive impairment, acting as a functional buffer even in the presence of Alzheimer's pathology or APOE4 risk alleles \citep{yang2024association, marselli2024protective, pettigrew2023alzheimer}. It is supported by specific neurobiological mechanisms, including greater functional connectivity in the Default Mode Network~\citep{steffener2011supporting, bartres2011structural}, neural compensation (recruiting bilateral regions per HAROLD/CRUNCH models) \citep{reuter2008neurocognitive, franzmeier2017cognitive}, and superior vascular resilience \citep{brenner2023cognitive}.

However, CR is not static. It is built or eroded by daily activity. Lifelong neuroplastic remodeling is central to maintaining this reserve \citep{gazerani2025neuroplastic, verkhratsky2024neuroglia}. This highlights the latent risk of the cognitive offloading and deskilling discussed in previous sections. If digital technologies replace the productive struggle of learning and reduce opportunities for complex problem-solving, they may systematically displace the stimulation required to build CR.

Emerging research supports this concern. Maladaptive technology use is already associated with stress, burnout, and depression \citep{kim2025ai, singh2025protecting, wiederhold2025cost, shanmugasundaram2023impact, paic2025impact}. If the automation of cognitive tasks leads to the atrophy of cortical networks, it effectively lowers the user's CR. This depletion could render individuals more vulnerable not only to cognitive decline but also to the physical pathologies that robust brains typically withstand \citep{fasoli2025dark, yan2024promises}. Thus, the preservation of analogue cognitive effort may be essential not just for skill retention, but for long-term physiological resilience \citep{meng2012education, gamble2025cognitive, wilson2019education}.
\section{Integrative Synthesis and Discussion}\label{sec:discussion}

The preceding sections mapped a highly fragmented literature into distinct cognitive domains (Section~\ref{sec:negative_effects_domains}) and unifying biological mechanisms (Section~\ref{sec:mechanisms}). In this section, we synthesize these findings into a cohesive theoretical framework. We highlight a fundamental shift in the landscape of cognitive risk—from resource allocation deficits driven by older technologies, to generative erosion driven by artificial intelligence. By examining the overarching moderators of these harms and the resulting efficiency-atrophy paradox, we provide a unified perspective on how digital technology reshapes human cognition, concluding with targeted calls to action for researchers, designers, and policymakers.

\subsection{Philosophical Context}\label{sec:philosophy}

While the preceding sections detail the empirical evidence of cognitive impairment, philosophical inquiry offers a theoretical framework for understanding these shifts. Philosophy has long examined technology’s effects on human cognition, debating whether tools extend our mental capabilities or cause them to atrophy.

Historically, the Extended Mind thesis posited that digital tools could validly expand human cognition, acting as external coupled systems that function analogously to biological memory \citep{clark1998extended}. However, phenomenologists and philosophers of technology have countered that the specific structure of digital tools inevitably reshapes cognitive habits for the worse. Early critiques argued that the rapid app-switching and superficial engagement promoted by smartphones would lead to attention fragmentation and the erosion of deep memory \citep{carr2010shallows, turkle2011alone, heidegger1977question}. More recently, philosophers of the ``attention economy'' have escalated this warning, arguing that systemic fragmentation does not merely distract, but fundamentally undermines the human will and the capacity to set independent goals \citep{williams2018stand}. This theoretical stance aligns closely with the empirical findings on media multitasking and inhibitory control discussed in Section \ref{sec:negative_effects_domains}.

Beyond basic cognition, philosophical discourse deepens the understanding of the epistemic and executive risks posed by AI. The concept of cognitive offloading is mirrored in warnings that overreliance on automation undermines expertise and alters perception \citep{dreyfus2008internet, ihde1990technology}. In the context of Generative AI, recent analyses characterize algorithmic outputs as philosophical ``bullshit''—speech generated without regard for truth, which risks degrading the user's own epistemic standards and care for verification \citep{hicks2024chatgpt}. As users increasingly rely on these outputs, the empirical reality of automation bias finds its theoretical root in the surrender of epistemic responsibility.

These concerns culminate in questions regarding the nature of human agency and ethical development. As algorithmic systems assume more decision-making power, scholars argue that human ethical reflection atrophies. Vallor describes this as a mirror effect, where AI traps users in a loop of recycled past data, flattening moral growth and preventing the cultivation of new virtues \citep{vallor2024mirror, zuboff2019age}. Generative humanism further argues that by reducing individuals from creators to mere transmitters of homogenized outputs, algorithms constrain freedom of choice \citep{brusseau2023mapping, clark2025extending}. Ultimately, philosophers and empiricists converge on a critical warning: that short-term gains in productivity may come at the cost of long-term cognitive development, limiting the human capacity for independent reasoning and ethical judgment \citep{siemens2022human, singh2025protecting, poszler2024impact}.

\subsection{The Distinct Profile of the Emerging AI Literature}

Divergences between the emerging AI literature and the broader digital media corpus largely reflect the former's nascent state. AI-focused research predominantly targets adult populations and higher-order domains such as critical thinking and creativity, distinct from the attention- and youth-focused trends in general screen-time research. Methodologically, the AI literature remains constrained by small-scale, cross-sectional, or single-session designs that rely heavily on self-reports and behavioral assessments. Furthermore, the ratio of theoretical to empirical contributions is notably higher in the AI domain, highlighting a gap in experimental validation.

\subsection{The Shift from Resource Allocation to Generative Erosion}

A distinct progression is evident in the type of cognitive impairment associated with different technological waves. The main group of literature (focusing on smartphones, multitasking, and the internet) predominantly describes deficits in cognitive resource allocation. Specifically, the fragmentation of attention, bottlenecks in working memory, and failures in inhibitory control. In contrast, the emerging AI literature signals a shift toward the erosion of generative and evaluative capacities. The risk has evolved from a difficulty in maintaining focus to a potential degradation of the ability to initiate, structure, and verify complex thought processes independent of algorithmic assistance. We are observing a transition from a crisis of attention to a crisis of agency.

\subsection{The Efficiency-Atrophy Paradox}

A consistent trade-off emerges across all cognitive domains: technologies that optimize short-term performance often undermine long-term competence. This efficiency paradox is observed in memory (GPS and search engines improve retrieval speed but weaken retention), learning (AI tutors accelerate task completion but bypass the productive struggle necessary for skill consolidation), and creativity (generative tools increase fluency but decrease diversity). The underlying mechanism appears to be the brain’s tendency to conserve metabolic energy by offloading demands to external systems. When this offloading becomes habitual rather than strategic, the neural substrates required for unassisted processing suffer from disuse atrophy.

\subsection{Universal Moderators of Harm}

\subsubsection{Passivity as the Primary Moderator of Harm}

The literature suggests that digital technology is not inherently deleterious. Instead, the magnitude of negative effects is strongly moderated by the mode of engagement. Harm is consistently linked to passive, unguided, or unregulated consumption, whether it is an infant passively watching a screen or a student copy-pasting an AI response. Conversely, scenarios involving active user interrogation, social co-viewing, or scaffolded use often mitigate these deficits. The critical distinction lies in whether the technology is used to substitute human cognition or to scaffold it.

\subsubsection{Displacement as a Universal Mechanism}

While specific neurobiological mechanisms vary, displacement acts as a universal driver of impairment across the lifespan. In early childhood, screens displace the sensory-motor interactions and caregiver reciprocity required for neurodevelopment. In adolescence and adulthood, digital consumption displaces the micro-rests and sleep required for memory consolidation and emotional regulation. In the context of AI, automated outputs displace the iterative cognitive processes (drafting, debugging, brainstorming) that constitute deep learning. Thus, a significant portion of the observed cognitive decline is not a direct result of toxic technological input but an indirect result of the loss of essential biological and cognitive maintenance activities.

\subsubsection{Systemic Feedback Loops}

The evidence points to the existence of recursive, self-reinforcing cycles where cognitive impairment begets further technological dependence. Neural adaptations such as dopaminergic tolerance create a physiological drive for increased stimulation, while the atrophy of executive control weakens the user's ability to regulate that consumption. Furthermore, as established in the review of cognitive epidemiology, these cognitive deficits may eventually compromise physical health and cognitive reserve, rendering the brain more susceptible to aging and pathology. This suggests that digital overuse is not merely a transient behavioral issue but a systemic risk factor that can alter the long-term cognitive and physiological trajectory of the user.

\subsubsection{Developmental Vulnerability and the Skill-Acquisition Gap}

The literature reveals a distinct developmental gradient regarding severity: the younger the brain, the more profound the potential structural and functional consequences. This heightened susceptibility in children and adolescents stems from two primary biological factors. First, neuroplasticity allows for rapid learning, but also makes the developing brain uniquely sensitive to environmental shaping. Excessive exposure to fragmented, high-speed digital stimuli during critical windows of myelination and synaptic pruning appears to wire the brain for distractibility rather than deep focus.

Second, there is a temporal mismatch in brain maturation. The limbic system (responsible for reward processing and dopamine response) matures earlier than the Prefrontal Cortex (responsible for impulse control and long-term planning). Digital interfaces designed with variable rewards and infinite scrolls effectively exploit this developmental gap, targeting the hyper-active reward centers of adolescents who biologically lack the fully developed brakes to regulate their engagement.

Finally, a crucial distinction emerges between adult and youth populations regarding AI. For adults, the risk is primarily the loss of previously mastered cognitive skills through disuse. For youth, the risk is a failure of acquisition. If generative AI substitutes for the productive struggle during formal education, students may never develop the foundational neural architectures for these capabilities in the first place. Thus, while adults risk becoming cognitively out of shape, the next generation risks a failure to launch fundamental critical thinking and creative capacities.

\subsection{Calls to Action}

\subsubsection{Research and Methodology}
Addressing the limitations of research into the effects of technology requires a methodological shift across the entire field of digital cognition. There is a broad consensus on the urgent need for more rigorous~\citep{beaudoin2024systematic, poupard2025systematic, yuce2025cost}, interdisciplinary~\citep{barros2024understanding, irzalinda2023screen, schacter2022media}, and longitudinal research~\citep{wang2023status, kirjakovski2023rethinking, moshel2024neuropsychological}. Scholars emphasize that understanding the complex effects of digital technologies—from smartphones to generative AI—requires moving beyond short-term, siloed investigations toward collaborative frameworks spanning psychology, neuroscience, education, ethics, and computer science.

\subsubsection{Human-Computer Interaction (HCI) and System Design}
Broad consensus suggests that while digital technologies pose cognitive risks, these can be mitigated through balanced use and intentional design. The literature emphasizes that AI should augment rather than replace cognition, with interfaces designed to scaffold attention and support deliberate allocation \citep{amalakanti2024impact, cualinescu2024impact, ali2024understanding, goldon2024short, clemente2024digital, korres2024unsupervised, schuster2025attention, bottger2023does, seitz2024impact}. Experts advocate for process-oriented systems that introduce desirable friction, Socratic questioning, and reflection prompts to maintain user autonomy and metacognition \citep{poszler2024impact, tankelevitch2025tools, hou2025appropriate}. Ethical design must prioritize human dignity and reasoning over efficiency, potentially reintroducing analogue strategies (for example, handwriting) to counterbalance digital influence \citep{alemdag2025effect, farkavs2024transforming, goyal2025ai, butler2024critical, manwell2022digital, joint2011if}.

\subsubsection{Education and Pedagogy}
In educational contexts, recommendations focus on structured, purposeful engagement guided by educators \citep{clemente2024digital, kostic2022digital, bezgodova2020smartphone}. To address integrity and bias, scholars propose AI tutors aligned with cognitive science that support desirable difficulties, alongside robust safeguards against plagiarism \citep{favero2025ai, bauer2025looking, shafei2025critical, sardi2025generative, Melisa2025CriticalThinkingAI, holzner2025generativeaicreativitysystematic}. Crucially, curricula must embed AI and digital literacy training for both students and teachers, while preserving in-person engagement and physical play to support social-emotional development \citep{kim2025ai, bai2023chatgpt, gong2024google, wiederhold2025cost, chen2024artificial}.

\subsubsection{Policy, Governance, and Institutional Frameworks}
Governance frameworks are essential to ensure privacy, transparency, and accountability. Drawing on international standards (for example, EU AI Act, AI4People), scholars call for policies that define acceptable use and uphold fairness \citep{paic2025impact, fasoli2025dark, duenas2024risks, kim2025ai, wang2025effect, dubey2024redefining, samala2025unveiling, yan2024promises, Eckhardt2024}. Institutions are urged to create sector-specific codes of conduct and integrate critical thinking into regulatory approaches, ensuring that cognitive protection is codified at the policy level \citep{gerlich2024balancing, rothinam2025systematic, leon2024potential}.

\subsubsection{Public Health and Societal Awareness}
Finally, public health strategies and awareness campaigns are needed to address misinformation, model collapse, and the trivialization of knowledge \citep{amalakanti2024impact, paulus2023screen, gong2024google, kim2025ai, fasoli2025dark, peschl2024human}. A society-in-the-loop approach emphasizes collective responsibility for cognitive well-being, supported by transparent AI tools and widespread training in recognizing manipulative design patterns \citep{george2024erosion, siemens2022human, skulmowski2024placebo, zhai2024effects, roussos2023mind, shanmugasundaram2023impact, chrzanowski2022changing, dergaa2024tools}.

\section{Conclusion}\label{sec:conclusion}

This review distinguishes itself from prior syntheses through its scope and integrative approach. While foundational reviews have isolated specific technologies or restricted their focus to isolated cognitive domains, we provide a systematic bridge between the established literature and the nascent body of work on AI. Furthermore, by applying principles of cognitive epidemiology, our analysis contextualizes these risks as potentially relevant factors for long-term health and longevity. The picture that emerges is consistent with a trade-off: as digital tools optimize the speed and efficiency of information processing, they may simultaneously disengage the cognitive effort hypothesized to support the neural substrates associated with deep, unassisted cognition.

Our analysis indicates a possible evolution in the landscape of these risks. Research on screens, smartphones, and the internet has predominantly characterized the challenge as one of resource allocation, particularly involving the fragmentation of attention and the bottlenecking of working memory. While the empirical evidence for AI is still in its infancy, early signals have been interpreted as suggesting a potential transition toward what may be described as generative erosion. By outsourcing the core processes of planning, writing, and reasoning, users may experience reduced engagement of the higher-order capabilities these systems emulate. The risk is no longer merely that we are too distracted to think deeply, but that sustained reliance may reduce opportunities for the habitual practice and exercise required to do so independently.

Mechanistically, these effects have been hypothesized to involve functional interference, neurochemical dysregulation, structural neuroplasticity, and psychosocial displacement. This latter mechanism may help explain longer-term associations, as digital engagement may displace essential biological and social activities. However, it is essential to emphasize that these associations are not purely technological. Instead, they are deeply intertwined with socioeconomic and environmental factors. As demonstrated in several longitudinal cohorts, technology use frequently acts as a marker for a lack of cognitive stimulation or reduced parental scaffolding rather than as a primary causal agent. If habitual digital offloading were to contribute to reductions in cognitive reserve, it might theoretically accelerate age-related decline.

Establishing definitive causality in long-term structural changes remains a significant challenge. While experimental studies confirm that digital interruptions acutely impair task performance, long-term structural changes are largely observed through correlational neuroimaging data. In these cases, the directionality between cognitive outcomes and heavy technology use remains difficult to isolate, as pre-existing cognitive or behavioral vulnerabilities often predict higher levels of technology consumption.

Ultimately, the literature indicates that these negative outcomes are not inevitable. Harm appears to be strongly moderated by the mode of engagement and individual self-regulation. Passive, unguided consumption has been associated with poorer outcomes, whereas scaffolded and intentional use may help preserve or potentially enhance human agency. The challenge for the future is therefore not to reject these technologies, but to design systems and pedagogical models that introduce desirable friction, with the aim of ensuring that digital tools serve as a scaffold for human intelligence rather than a total substitute for it.

\subsection{Limitations}

First, our primary objective was to identify and categorize documented risks and negative effects. Consequently, our search strategy and inclusion criteria were deliberately biased toward characterizing the upper bound of potential harm. This review should therefore be read as a mechanistic diagnostic of risk rather than a comprehensive cost-benefit analysis of digital integration.

Second, the two bodies of literature analyzed in this review differ significantly in maturity. The findings regarding screens, smartphones, and multitasking are supported by over a decade of longitudinal and experimental research, offering high confidence in the observed effects. In contrast, the literature on AI is nascent and dominated by cross-sectional, single-session studies and theoretical papers. The hypothesized shift in risk profile identified in this paper must therefore be viewed as a preliminary framework that requires urgent longitudinal validation.

Third, while we employed a semi-automated snowball search to maximize coverage, our reliance on specific seed papers means we may have missed relevant studies. In addition, because citation expansion relied on bibliographic linkages indexed within the OpenAlex database, relevant publications may have been omitted if they were incompletely indexed or not bibliographically connected within this dataset. However, if such studies exist, they are likely bibliographically disconnected from the main clusters of psychology and human–computer interaction research captured by the database at the time of retrieval.

\section*{Acknowledgements}

The authors would like to thank Grega Repovš for his comments and suggestions.

\section*{Declaration of competing interest}

The authors declare that they have no known competing financial interests or personal relationships that could have appeared to influence the work reported in this paper.

\section*{Funding}

This work was supported by the Slovenian Research Agency (Grants No. P2-0426, P2-0442, J5-60084, and J6-60105).

\section*{Declaration of generative AI and AI-assisted technologies}

During the preparation of this work the authors used ChatGPT 5.x and Gemini Pro 3.x for language refinement and proofreading. After using this tool/service, the authors reviewed and edited the content as needed and take full responsibility for the content of the published article.

\bibliographystyle{apalike}
\bibliography{bibliography}

\appendix

\section{Related Review Papers}\label{sec:appendix_related_reviews}

\begin{table*}[ht]
\centering
\captionof{table}{Related review papers, their thematic focus, and the number of studies included (\#). When the number of studies was not explicitly reported, it was estimated from the paper and is indicated with an asterisk (*):}
\label{tbl:appendix_related_reviews}
\begin{scriptsize}
\begin{tabular}{p{3.5cm}p{0.2cm}p{8.8cm}}
\toprule
Paper & \# & Focuses on the effects of \\
\midrule
\cite{anna2025extended} & 29* & LLMs on critical thinking, and problem-solving skills \\
\cite{adams2023screen} & 10 & screen time on executive functions, working memory, inhibition, attention, cognitive flexibility, and social contingency in infants \\
\cite{alamri2023relationship} &  19 & smart media on language development in children \\
\cite{barros2024understanding} & 22* & digital technology on cognition in the context of neuroplasticity \\
\cite{beaudoin2024systematic} & 55* & lifelong technology experience on creativity, adaptability, and decision-making \\
\cite{bulut2023association} &22* & internet and social media use on attention and media multitasking in adolescents and young adults \\
Clemente-Su{\'a}rez...\citeyearpar{clemente2024digital} & 157 & digital device usage on attention, memory, executive functions, problem-solving skills, and social cognition in children \\
Deckker...\citeyearpar{deckker2025systematic} &  70* & AI on memory, attention, decision-making, and social cognition \\
\cite{dubey2024redefining} & 29 & ChatGPT on attention, executive function, language, memory, visuospatial functions \\
\cite{firth2019online} & 90* & internet use on attention, transactive memory, social cognition, and creativity in children, adolescents, and older populations \\
\cite{ge2025aigc} & 41 & how AI generated content impact critical thinking in designers \\

\cite{george2024erosion} & 18* & technology on critical thinking, problem-solving, and creativity \\
\cite{goyal2025ai} & 18 & AI systems on critical thinking in students \\
\cite{holzner2025generativeaicreativitysystematic} & 28 & generative AI on creativity \\

\cite{jourdren2023impact}  & 15 & screen time on attention in children \\
\cite{kundu2024psychological} &  24 & AI on cognition, emotions, and behavior in children and adolescents \\
\cite{liu2022screen} & 60* & screen media overuse on sleep, obesity, emotional and behavioral problems in children and adolescents \\
Mallawaarachchi...\citeyearpar{mallawaarachchi2024early}  & 64 & screen time on executive functions, language, and academic performance in children \\
\cite{marciano2021developing} & 16 & screen-based media use on cognitive control and addictive behaviors in adolescents \\
\cite{massaroni2023relationship} & 18 & screen time on language development in children \\
\cite{marsh2022digital} &  194 & digital workplace on technostress, overload, addiction, anxiety, interruption and distraction, work-nonwork issues \\
\cite{Melisa2025CriticalThinkingAI} & 19 & ChatGPT on critical thinking, evaluation, and independent judgment skills in students \\
\cite{moongela2024effect} & 44 & ChatGPT on cognitive thinking skills in students \\
\cite{paulus2023screen} & 37 & screen media activity on sleep patterns, mood disturbances, anxiety, and cognition in teenagers \\
\cite{Premkumar2024Impact} & 30 & generative AI on critical thinking skills in students \\
\cite{qu2025generative} & 18 & Generative AI tools on lower-order cognitive skills, and higher-order cognitive skills in students \\
\cite{santos2022association} & 11 & screen time on attention in children \\
\cite{sardi2025generative} & 38 & generative AI on learning and critical thinking skills in students \\
Shanmugasundaram...\citeyearpar{shanmugasundaram2023impact} & 80* & digital technology, social media, and AI on attention, memory, addiction, novelty-seeking, perception, decision-making, critical thinking, and learning \\
\cite{Thaker2025DualImpactAI} & 15 & AI use on learning, memory, creativity, and decision-making \\
Vedechkina...\citeyearpar{Vedechkina2021} &  80* & digital technology on attention in children \\
\cite{wang2025effect} & 51 & ChatGPT on learning and higher-order thinking in students \\
Wiradhany...\citeyearpar{wiradhany2021everyday} & 16 & multitasking on attention, memory, inhibition, and behavior  \\
\cite{wu2024early} & 33 & screen-based media use on executive functions in children \\
\cite{zhai2024effects} & 14 & AI on decision-making, critical thinking, analytical thinking, and over-reliance in students \\
\end{tabular}
\end{scriptsize}
\end{table*}

See Table~\ref{tbl:appendix_related_reviews}.

\section{Detailed Scoping Review Methodology} \label{sec:appendix_detailed_meth}

\subsection{Literature Search - Main Paper Set}\label{sec:main_search}

The goal of this literature search was to find empirical (primary goal) and non-empirical (secondary goal) papers that focus on the effects of technology on human cognition. Furthermore, to be included, a paper must at least partially focus on negative effects.

Searching for papers on the negative effects of technology on human cognition is challenging for several reasons. First, the relevant body of literature on the effects of technology is large. Second, the relevant body of literature is diverse in field (health, psychology, education, economics), type of technology, and part of cognition that is affected, making keyword search ineffective. And third, it is difficult to quickly identify papers that at least partially focus on negative effects.

To address these challenges, we perform a semi-automated snowball search using the OpenAlex \citep{Priem2022} index of academic works. At each iteration, we use automated keyword-based filtering to limit the papers to a feasible subset, followed by manual inspection and filtering. We iterated the snowball search until no new papers were added. We provide a more detailed description in the remainder of this section.

The process described below is summarized in the flow diagram in Figure~\ref{fig:prisma_main}.

\subsubsection*{Step 1: Seed Papers}

As a starting point for the snowball search, we used review papers on the effects of technology on human cognition. Review papers are typically hubs in the bibliographic citation graph (a large number of references and citations), and it is unlikely that there is no citation path between a relevant paper and the review papers in its field.

We selected the review papers by first performing a keyword search on Scopus (July 23, 2024) using this query string:

\begin{scriptsize}
\begin{verbatim}
TITLE (

    ("survey" OR "review" OR "meta-analysis" OR "overview") AND
    
    ("cognitive" OR "cognition" OR "attention" OR "memory" OR 
     "critical thinking" OR "decision making") AND 
     
    ("technolog*" OR "intelligen*" OR "AI" OR "software" OR
     "comput*" OR "information retrieval" OR "search*" OR "google")
     
)
\end{verbatim}
\end{scriptsize}

The three parts of the query aim at identifying review papers, papers focusing on cognition, and papers focusing on technology. Note that \emph{perception} is a common part of cognition, but it is also a commonly used term in paper titles in other fields, so we did not include it. Furthermore, it was necessary to limit the search to paper titles; otherwise, the number of results would be prohibitive for manual filtering.

The query returned 938 papers, which we manually filtered with the inclusion criteria \emph{a review focusing on the impact of technology use on cognition}. The final result were 13 papers \citep{Baltes2002156,Benzinger2023,Kristóf2016127,Vedechkina2021,Wang2023301,zhai2024effects,adams2023screen,beaudoin2024systematic,Ciriello2023,gong2024google,poszler2024impact,poupard2025systematic,wilmer2017smartphones}.

\subsubsection*{Step 2: First Iteration of Snowball Search}

The 13 seed papers were cited by 804 papers (forward snowball) and referenced 1150 papers (backward snowball). After filtering with the criterion \emph{focusing on the negative impact of technology use on cognition}, this was reduced to 76 papers in the forward and 169 in the backward list. Combined, there were 219 new and unique papers, 232 combined with the 13 seed papers.

\subsubsection*{Step 3: Remaining Iterations of Snowball Search}

From this point onward, we use a mostly automated snowball search process using the OpenAlex API. The 232 papers are used as seed papers for a repeated application of the following procedure:

\begin{enumerate}
    \item Retrieve from OpenAlex all papers that either reference or are referenced by a paper from the set of seed papers.
    \item Automatically filter papers to keep only articles, reviews, and book chapters.
    \item Automatically filter papers to keep only those that meet at least one of the two criteria:
    \begin{itemize}
        \item The paper has a bibliographic link to at least 10 papers in the seed set. This ensures that any at least moderately connected paper is manually inspected.
        \item The paper has a bibliographic link to at least 3 papers in the seed set and matches a modified search string. The modified search string omits the line that identifies review papers and is extended to the title or abstract.
    \end{itemize}
    \item Automatically exclude papers that have already been manually inspected in previous iterations.
    \item Manually inspect remaining papers and filter papers to keep only those that are \emph{focusing on the negative impact of technology use on cognition}.
    \item Add remaining papers to the set of seed papers.
\end{enumerate}

\begin{table}[ht]
\centering
\caption{Paper counts for all snowball search iterations.}
\label{tab:example}
\begin{tabular}{lrrrrr}
\toprule
Iter & Seed & Snowball & Pass auto & Pass manual \\
\midrule
manual & 13 & 1150 & - & 219 \\
1 & 232 & 27577 & 1384 & 501 \\
2 & 733 & 54639 & 381 & 120 \\
3 & 853 & 58043 & 78 & 30 \\
4 & 883 & 59157 & 12 & 9 \\
5 & 892 & 59444 & 1 & 0 \\
\bottomrule
\end{tabular}
\end{table}

Note that the same work can appear multiple times under different OpenAlex IDs. For example, if a paper is published in a journal and in a public repository. After removing duplicates, we were left with 888 papers.

\subsubsection*{Step 4: Postprocessing}

The remaining 888 papers were then categorized (see \ref{sec:categorization}). During this process, we further identified 38 papers that did not focus on technology or papers that did not claim a clear negative impact. We were unable to access the full text of three papers~\citep{Murphy2019,sharmandemola2023effect, velasco2020impact}.

The final result of the process is a list of 563 empirical papers and a list of 284 non-empirical papers. Both are made available as supplementary material.

\begin{figure*}[ht]
    \centering
    \begin{tikzpicture}[
        node distance=1.5cm and 2cm,
        box/.style={rectangle, draw, rounded corners, align=center, minimum width=5cm, minimum height=1.2cm, text width=5.5cm, font=\small},
        exbox/.style={rectangle, draw, align=left, minimum width=4.5cm, text width=5.5cm, font=\small},
        arrow/.style={thick, ->, >=stealth}
    ]

    \node [box] (id) {\textbf{Identification}\\ Scopus Keyword Search (\textit{n} = 938)\\ \textit{Filtered to 13 seed papers}};
    
    \node [box, below=of id] (screen) {\textbf{Screening}\\ OpenAlex Snowballing (5 Iterations)\\ \textit{n} = 888 unique records retrieved};
    
    \node [box, below=of screen] (eligibility) {\textbf{Eligibility}\\ Full-text postprocessing \\ \textit{n} = 888};
    
    \node [exbox, right=of eligibility] (exclude) {\textbf{Records excluded (\textit{n} = 41):}\\ $\bullet$ No tech / no clear negative impact (\textit{n} = 38)\\ $\bullet$ Full text unretrievable (\textit{n} = 3)};
    
    \node [box, below=of eligibility] (included) {\textbf{Included (Main Set)}\\ \textbf{\textit{N} = 847}\\ Empirical: \textit{n} = 563\\ Non-empirical: \textit{n} = 284};

    \draw [arrow] (id) -- (screen);
    \draw [arrow] (screen) -- (eligibility);
    \draw [arrow] (eligibility) -- (exclude);
    \draw [arrow] (eligibility) -- (included);

    \end{tikzpicture}
    \caption{Flow diagram detailing the study selection process for the Main paper set.}
    \label{fig:prisma_main}
\end{figure*}

\vspace{1cm}

\begin{figure*}[ht]
    \centering
    \begin{tikzpicture}[
        node distance=1.5cm and 2cm,
        box/.style={rectangle, draw, rounded corners, align=center, minimum width=5cm, minimum height=1.2cm, text width=5.5cm, font=\small},
        exbox/.style={rectangle, draw, align=left, minimum width=4.5cm, text width=5.5cm, font=\small},
        arrow/.style={thick, ->, >=stealth}
    ]

    \node [box] (id) {\textbf{Identification}\\ Scopus Keyword Search (\textit{n} = 21)\\ \textit{Filtered to 7 + 8 known = 15 seeds}};
    
    \node [box, below=of id] (screen) {\textbf{Screening}\\ OpenAlex Snowballing (2 Iterations)\\ \textit{n} = 329 unique records retrieved};
    
    \node [box, below=of screen] (eligibility) {\textbf{Eligibility}\\ Full-text postprocessing \\ \textit{n} = 329};
    
    \node [exbox, right=of eligibility] (exclude) {\textbf{Records excluded (\textit{n} = 243):}\\ $\bullet$ Not AI / no clear negative impact\\ $\bullet$ Full text unretrievable (\textit{n} = 2)};
    
    \node [box, below=of eligibility] (round1) {\textbf{Initial Inclusion}\\ \textit{n} = 86\\ Empirical: \textit{n} = 21\\ Non-empirical: \textit{n} = 65};

    \node [box, right=of round1] (update) {\textbf{Additional Snowball Search}\\ (August 2025)\\ \textit{n} = 4 non-empirical papers};

    \node [box, below=of round1] (included) {\textbf{Included (AI Set)}\\ \textbf{\textit{N} = 90}\\ Empirical: \textit{n} = 21\\ Non-empirical: \textit{n} = 69};

    \draw [arrow] (id) -- (screen);
    \draw [arrow] (screen) -- (eligibility);
    \draw [arrow] (eligibility) -- (exclude);
    \draw [arrow] (eligibility) -- (round1);
    \draw [arrow] (update) -- (included);
    \draw [arrow] (round1) -- (included);

    \end{tikzpicture}
    \caption{Flow diagram detailing the study selection process for the AI paper set, including the supplementary snowball iteration.}
    \label{fig:prisma_ai}
\end{figure*}

\subsection{Literature Search - AI Paper Set}\label{sec:ai_search}

As in the search in Appendix \ref{sec:main_search}, the goal of this literature search was to find empirical (primary goal) and non-empirical (secondary goal) papers that at least partially focus on the negative effects of technology. Additionally, this search was limited to papers published in 2023 or later, where the technology involved is AI.

The additional temporal and technological constraint allows us to use a manual snowball search until no new papers are found. We provide a more detailed description in the remainder of this section.

The process described below is summarized in the flow diagram in Figure~\ref{fig:prisma_ai}.

\subsubsection*{Step 1: Seed papers}

We collected seed papers by first performing a keyword search on Scopus (June 26, 2025) using this query string:

\begin{scriptsize}
\begin{verbatim}
TITLE (
    "artificial intelligence" OR
    "AI" OR
    "LLM" OR
    "ChatGPT" OR
    "large language model*"
) AND 

TITLE-ABS-KEY (
    "cognitive debt" OR 
    "cognitive offloading" OR
    "cognitive reserve" OR 
    "digital dementia" OR 
    "digital amnesia"
) AND 

PUBYEAR >  2022
\end{verbatim}
\end{scriptsize}

The three parts of the query aim at identifying AI papers, papers focusing on cognition, and papers published in 2023 or later.

The query returned 21 papers, which we manually filtered to 7 papers that met the inclusion criteria \emph{focusing on the negative effects of AI use on cognition}. We combined these 7 papers with 8 additional relevant papers, for a total of 15 seed papers \citep{ccela2024risks,doshi2024generative,essien2024influence,gerlich2025ai,karny2024learning,kazemitabaar2023studying,kosmyna2025your,leon2024potential,naseer2025psychological,shu2024ai,shukla2025skilling,shum2024generative,vasconcelos2023explanations,vzorin2024chatgpt,wahid2025explainable}

\subsubsection*{Step 2: Snowball Search}

Seed papers were cited by 1076 papers (forward snowball) and referenced 648 papers (backward snowball). After filtering for inclusion criteria and removing duplicates, the remaining 58 papers were used in the second snowball iteration. These papers were cited in 4,051 papers and referenced in 4,314 papers. After filtering for inclusion criteria and removing duplicates, 256 papers remained, for a total of 329 papers. We terminated snowballing after two iterations, because the third iteration papers were either duplicates or published before 2023.

\subsubsection*{Step 3: Postprocessing}

The remaining 329 papers were then categorized (see \ref{sec:categorization}). During this process, we further identified 242 papers that did not focus on AI or papers that did not claim a clear negative impact. The main reason for this large discrepancy is papers that claim negative impact, but it is revealed upon closer inspection that the claims are not empirically justified and/or are a reference to past work, without a meaningful contribution. We were not able to access the full text of two papers~\citep{Kreijkes2026,liu2025university}. The results are 21 empirical papers and 65 non-empirical papers. 

\subsubsection*{Step 4: Additional Snowball Search}

Due to the rapid development of AI and increasing interest in the effects of AI on humans, we performed another, more recent (August 29, 2025) iteration of snowballing, with the 21 empirical papers as seeds. This resulted in 4 new non-empirical papers.

The final result of the process is a list of 21 empirical papers and a list of 69 non-empirical papers. Both were appended to the empirical and non-empirical paper lists that are provided as supplementary material.

\subsection{Categorizing Empirical Papers}\label{sec:categorization}

For every empirical paper in the Main and AI paper sets, we also include the following information:

\begin{itemize}
    \item Basic bibliographic data: \textbf{WID} (OpenAlex ID), \textbf{DOI}, \textbf{Publication title}, \textbf{Publication source}, \textbf{Publication year}.
    \item \textbf{Number of participants.} If there is more than one study, we record the number of participants for each.
    \item \textbf{Technology.} Type of technology being studied:
        \begin{itemize}
            \item AR / VR
            \item computers / laptops,
            \item digital devices / technologies,
            \item digital media,
            \item GPS / navigational aids,
            \item internet,
            \item multitasking,
            \item smartphones / mobile devices,            
            \item social media networks,
            \item television,
            \item texting / emails / messaging,
            \item video games / gaming,
            \item other (subgroup recorded in parentheses).
        \end{itemize}
    \item \textbf{Cognitive domain.}
        \begin{itemize}
            \item attention and concentration
            \item executive functioning
            \item language/verbal skills
            \item memory
            \item motor skills and construction
            \item perception
            \item processing speed
            \item sensation
            \item other (subgroup reported in parentheses).
        \end{itemize}
    \item \textbf{Age group.} Participant age, partitioned into 7 groups:
        \begin{itemize}
            \item infants and toddlers (0–3 years),
            \item preschoolers (3–5 years),
            \item children (6–12 years, includes school-aged children),
            \item adolescents (13–18 years, includes middle and high school students),
            \item young adults (18–29 years, includes university students),
            \item adults (30–64 years, includes general population and professionals), 
            \item older adults (65+ years).
        \end{itemize}
    \item \textbf{Time span.} Duration of the study:
        \begin{itemize}
            \item one-time,
            \item days,
            \item weeks,
            \item months,
            \item years.
        \end{itemize}    
    \item \textbf{Study type.}
        \begin{itemize}
            \item quasi experimental study,
            \item true experimental study,
            \item prospective cohort study,
            \item retrospective cohort study,
            \item cross-sectional study,
            \item case-control study,
            \item other (quantitative),
            \item other (qualitative).
        \end{itemize}    
    \item \textbf{Assessment methods.}
\end{itemize}

Note that if multiple values apply to a paper, we record all of them. For example, a paper can feature multiple studies or a study can utilize multiple methodologies or cover multiple groups of participants.

\section{Summary of Empirical Papers}\label{sec:empirical2}

In this section, we provide a broad overview of what is being researched in the area of the negative effects of technology use on cognition and how these studies are conducted. Our analysis is based on two groups of papers: (main group) 563 empirical papers identified through our main systematic search (see \ref{sec:main_search}) and (AI group) additional 21 recent papers focusing specifically on AI, identified through a separate search (see \ref{sec:ai_search}). Throughout this section, we compare these two groups of papers. Each paper was categorized according to publication source, year, technology, cognitive domain, target population age group, participant count, study duration, and study type (see \ref{sec:categorization}). The complete list of papers, along with their associated metadata, is provided as supplementary material.

Figure 2 in the paper presents the statistical summaries. Most categorical variables are summarized as the proportion of papers featuring that category, accompanied by an interval quantifying the uncertainty in that estimate. All reported intervals are Bayesian credible intervals obtained using a Binomial-Beta model with a Beta($\frac{1}{2}, \frac{1}{2}$) prior (Jeffreys’ prior) for the proportion. Note that all proportions are \textit{per paper}, but a single paper can feature more than one category. Consequently, the proportions may sum to more than one within a given category.

Publication year results indicate that the number of papers per year is increasing, a trend observed in both the main and AI-focused groups. Because overall research output is also growing, it is difficult to determine whether papers on this topic are increasing disproportionately. Although papers focusing on AI existed prior to 2022, their number was relatively small compared to the past three years, with only 4 such papers appearing in the main group.

The distribution of publication sources exhibits a long tail, indicating that many venues are appropriate for research on the negative effects of technology on cognition, yet no single journal is dedicated exclusively to this topic. This observation is in line with our view that the field is broad and fragmented. Some journals, however, stand out due to a disproportionately high number of relevant papers, notably Computers in Human Behavior and Computers \& Education. The high publication counts in mega-journals such as PLOS ONE and Scientific Reports, and to some extent Frontiers in Psychology, should be interpreted cautiously given the overall larger number of papers these journals publish.

\begin{figure*}[t]
    \centering
    \includegraphics[width=0.75\linewidth]{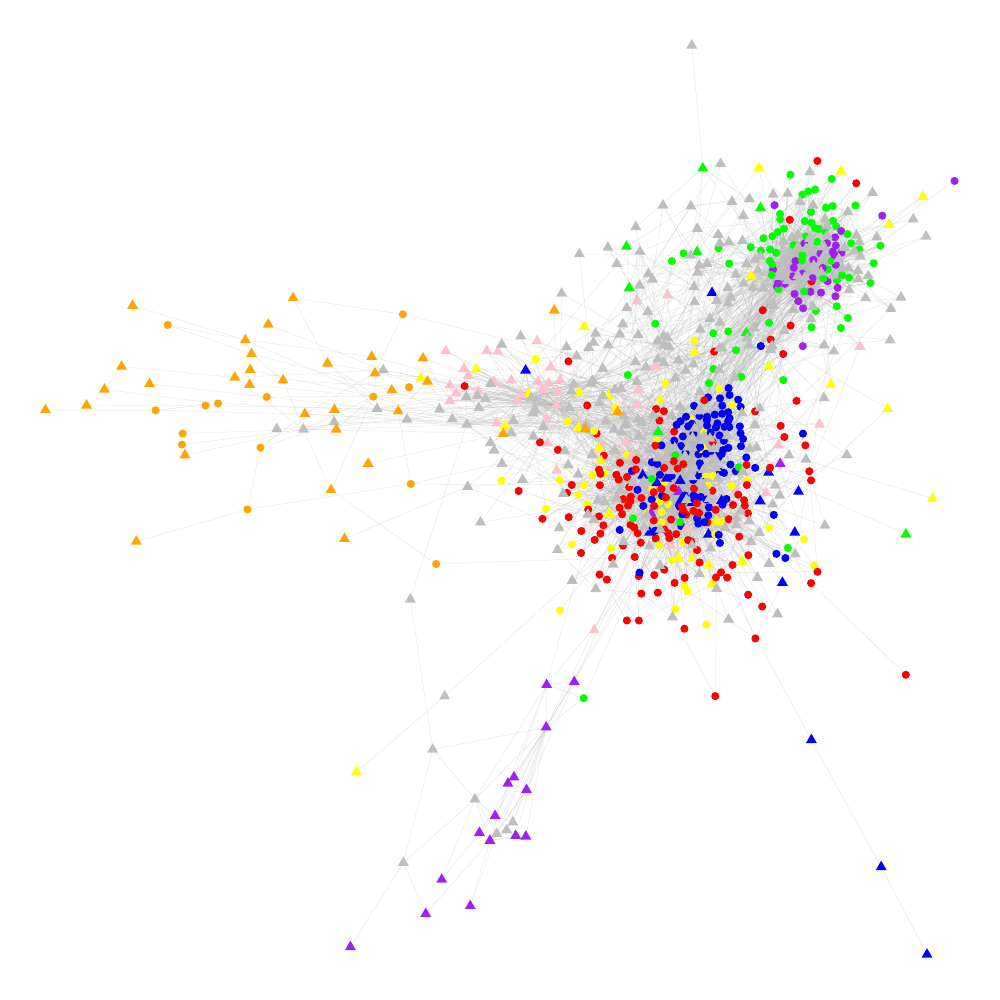}
    \caption{Citation network for the empirical papers with non-empirical papers included. Papers with 0 degree are excluded. The layout is the result of the Fruchterman-Reingold algorithm. An empirical paper's color is determined by our categorization of technology: smartphones/mobile devices (\textcolor{red}{\textbullet}), 
multitasking (\textcolor{blue}{\textbullet}), 
digital media (\textcolor{green}{\textbullet}), 
social media networks (\textcolor{yellow}{\textbullet}), 
internet (\textcolor{pink}{$\blacktriangle$}), 
television (\textcolor{Purple}{\textbullet}), 
digital devices/technologies (\textcolor{yellow}{$\blacktriangle$}), 
video games/gaming (\textcolor{green}{$\blacktriangle$}), 
ar/vr (\textcolor{Purple}{$\blacktriangle$}), 
texting/emails/messaging (\textcolor{blue}{$\blacktriangle$}), 
computers/laptops (\textcolor{red}{$\blacktriangle$}), and other (AI) (\textcolor{orange}{\textbullet}). Non-empirical papers from main group (\textcolor{gray}{$\blacktriangle$}) and from AI group (\textcolor{orange}{$\blacktriangle$}).}
    \label{fig:network}
\end{figure*}

The most common technologies of focus are smartphones and other mobile devices, which appear in roughly 20–30\% of papers. These are followed by work with a more specific emphasis on multitasking and a broader focus on digital media. Rounding out the top five are studies on social media networks and on the internet more generally. Together, these categories account for the majority of papers.

Figure~\ref{fig:network} shows a citation network of the papers, colored by technology. Our choice of categories is largely validated, as papers within the same category tend to cluster together. This pattern also aligns with our view that the field remains broad and fragmented.

Research on smartphones, multitasking, social media networks, and digital devices and technologies appears to be relatively well-connected. Work with a more general focus on the internet is somewhat separate from this main cluster, and research on AR/VR is even more distinct. Research on digital media and on television is also mostly separate from the main cluster, but the two are closely connected to each other, suggesting that they may constitute a single community cluster.

\begin{figure*}[t]
    \centering
    \includegraphics[width=0.6\linewidth]{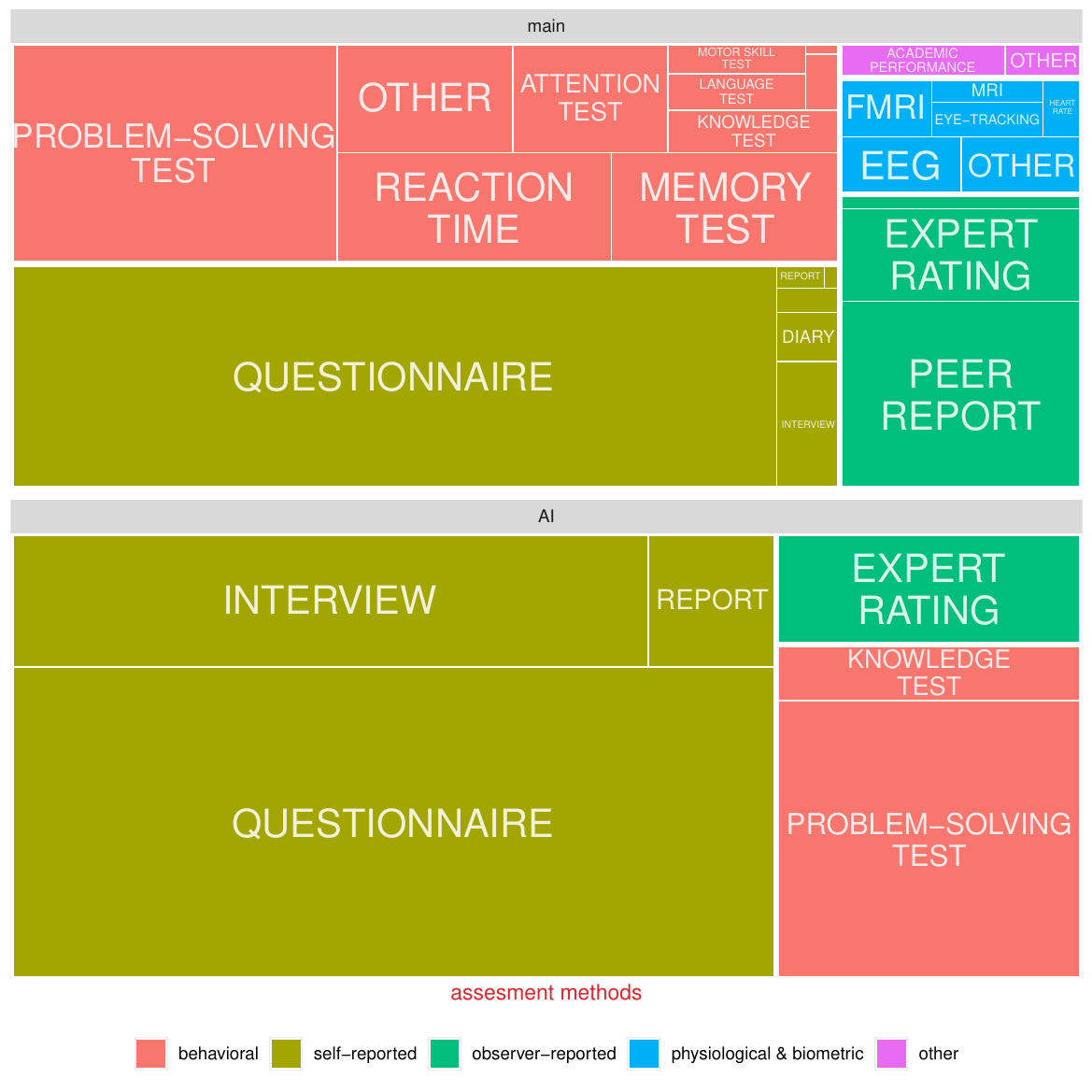}
    \caption{A summary and comparison of the prevalence of different assessment methods in empirical studies.}
    \label{fig:methodology}
\end{figure*}

Most empirical papers in the AI group have no citation links to papers in the main group. In other words, recent AI-focused work does not cite relevant empirical findings from related fields, at least not directly. It is possible that these findings are incorporated indirectly, through summaries provided in non-empirical papers, although most non-empirical papers from the AI group are also disconnected from the main group. The empirical AI papers also, with a few exceptions, do not cite each other. This is to be expected, given the relatively short time span in which they were produced (most of the papers can be considered concurrent works). It is also worth noting that there are disproportionately as many non-empirical papers in the AI group as there are empirical papers.

In the main group, roughly 35\% of the papers focus on attention, and just under 20\% each examine memory and executive functioning. These three basic cognitive domains therefore account for the majority of research, with academic performance being the only higher-order cognitive domain in the top five, at approximately 20\% of papers. This stands in stark contrast to the AI group, where research centers on higher-order cognitive domains such as critical thinking, creativity, specific task performance, and mental health. AI group papers also appear to focus somewhat more on adults and less on adolescents, with little to no work on younger populations, which aligns with the demographics of current AI users.

Most studies include around 100 participants, with the distribution fairly symmetrical when viewed on the log scale. The sample sizes of recent AI papers are too small to conclude that their distribution differs from that of the main group. Large population-based or longitudinal studies also cannot reasonably be expected this early in the investigation of a new technology. It is therefore unsurprising that all AI group studies are single-time studies, with only two also including components lasting on the order of weeks. Single-time studies also account for roughly 75\% of studies in the main group, although longer-term designs are well represented there.

Most AI studies are cross-sectional, with the remainder being experimental. These are also the two most common study types in the main group, which includes a proportionally larger share of experimental studies and additional designs, particularly prospective cohort studies. Again, this discrepancy may simply reflect the fact that research on the effects of AI is still at a relatively early stage.

Figure~\ref{fig:methodology} shows that both the main group and the AI group rely most heavily on self-reported, behavioral, and observer-reported assessments, in that order. However, AI group papers use self-reported assessments disproportionately more. Moreover, physiological, biometric, and other forms of assessment are almost absent in the AI group. For now, this may also be attributed to the relatively early stage of AI-related research, as these assessment types are less common in the main group as well.

\end{document}